\documentclass[a4paper,12pt]{article}
\RequirePackage[fleqn]{amsmath}
\usepackage{url}
\usepackage{graphicx}
\usepackage{amssymb}
\usepackage{undertilde}
\usepackage{booktabs}
\usepackage{xcolor,colortbl}
\usepackage[latin1]{inputenc}
\usepackage{rotating}
\usepackage{tikz}
\usepackage{textcomp}
\usepackage{subcaption}

\usepackage{makeidx}         % allows index generation
\usepackage{graphicx}        % standard LaTeX graphics tool
\usepackage{multicol}        % used for the two-column index
\usepackage[bottom]{footmisc}% places footnotes at page bottom
\RequirePackage[OT1]{fontenc}
\RequirePackage[fleqn]{amsmath}
\usepackage{graphicx}
\usepackage{amssymb}
\usepackage{booktabs}
\usepackage{xcolor,colortbl}
\usepackage[latin1]{inputenc}
\usepackage{makeidx}         % allows index generation
\usepackage{graphicx}        % standard LaTeX graphics tool
\usepackage{multicol}        % used for the two-column index
\usepackage[bottom]{footmisc}% places footnotes at page bottom
\usepackage{natbib}
\usepackage{caption} % per caption tabelle
\usepackage{multirow} % per  tabelle
\usepackage[normalem]{ulem} % per correzioni
\usepackage{rotating}

\usepackage{setspace}
\usepackage[font=small]{caption}
\usepackage{breakurl}

\begin{document}

\newcommand{\tr}{^{\prime}}

\def\b#1{\mbox{\boldmath $#1$}}    % - \b: grassetto in formula
\def\cg#1{\ensuremath{\mathcal{#1}}}      % caratteri calligrafici
\def\cgl#1{\mbox{\scriptsize {${\cal #1}$}}}

\renewcommand{\baselinestretch}{1.2}

\begin{center}
{\Large\ \textbf{Hidden Markov Models for Longitudinal Rating Data\\
		\vspace{0.2 cm} with Dynamic Response Styles}
 } \

\vspace{0.6 cm}
%\vspace{0.6 cm}

\text{Roberto Colombi}$^{*}$, \text{Sabrina Giordano}$^{**}, $ \text{Maria Kateri}$^{***}$ \\  \ \\
\small{
	\textit{$^{*}$ Department of Management, Information and Production Engineering, University of Bergamo \\ e-mail: roberto.colombi@unibg.it}\\[0.5 cm]
	\textit{$^{**}$ Department of Economics, Statistics and Finance ``Giovanni Anania", University of Calabria \\ e-mail: sabrina.giordano@unical.it}\\[0.5 cm]
 \textit{$^{***}$ Institute for Statistics, RWTH Aachen University \\ e-mail:
maria.kateri@rwth-aachen.de}
}
\end{center}
\vspace{0.2 cm}
\abstract{	
	This work deals with the analysis of longitudinal ordinal responses. The novelty of the proposed approach is in
modeling simultaneously
the temporal dynamics of a latent trait of interest, measured via the observed ordinal responses, and the answering behaviors influenced by response styles, through hidden Markov models (HMMs) with two latent components.
This approach enables the modeling of (i) the substantive latent trait, controlling for response styles;
 (ii) the change over time of latent trait and answering behavior, allowing also dependence on
individual characteristics. For the proposed HMMs, estimation procedures,  methods for standard errors calculation, measures of goodness of fit and classification, and full-conditional residuals are discussed. The proposed model is fitted to ordinal
longitudinal data from the Survey on Household Income and Wealth (Bank of Italy) to give insights on the evolution of the Italian households financial capability.}

\vspace{0.3 cm}


% Produce the title:

\doublespacing

%\maketitle

%***********************************************************************

%%%%%%%%%%%%%%%%%%%%%%%%%%%%%%%%%%%%%%%%%%%%

\section{Introduction}\label{Sec.Intro}

Psychometric literature widely debated the different behavior patterns of respondents to rating surveys, which may introduce distortions or inaccuracies in their responses. Questions on attitudes, opinions, perceptions are usually Likert-type or rating-scale
items, and the observed responses may not reflect the respondents' true preferences but their tendency to use only a small number of the available rating scale options, governed by an underlying behavioral mechanism, known as Response Style (RS) \citep[e.g.,][for an overview]{van2013response}.
The activation of a
response style mechanism
influences systematically the way interviewees use response scales, introducing bias in the responses and
scale usage heterogeneity, which may impact the data quality and  the validity of the results
\citep[e.g.,][]{baumgartner2001, roberts2016}. %Several RSs
%have been identified and studied \citep[e.g.,][for an overview]{van2013response}.
%In this paper, a model is introduced that enables capturing the most commonly encountered RS such as
%tendency to select at random categories (careless RS, CRS),
%tendency to prefer positive response categories/answer with agreement (acquiescent RS, ARS),
%or negative response categories (disacquiescent RS, DRS), middle/neutral categories (middle
%RS, MRS), or extreme categories (extreme RS, ERS).
%, that capture the major biases in questionnaire responses

What is new in our approach
is the interest on the longitudinal perspective where respondents are asked, at several time occasions, to give a
subjective assessment about rating-scale items and their responses are indicators of a latent trait of interest (e.g., health status,
environmental risk, customer satisfaction). Moreover, responses can be driven or not by RS, the RS attitude can vary dynamically and the change over time of responses and answering behaviors can depend also on individual characteristics.

More precisely, in the context of longitudinal ordered categorical data analysis, the  methodological  contribution of the paper is a hidden Markov model (HMM) with a bivariate latent Markov chain that jointly models an unobservable trait of interest
and an unobservable binary indicator of the respondent's form of answering (response style driven or not) over time.  The use of HMMs in the context of categorical longitudinal data is not new \citep[see][for a comprehensive review]{bartolucci2012latent}, but to date  there does not exist any HMM-based procedure useful for modeling the evolution of an underlying response behavior  over time.
A further contribution of the proposed approach lies on providing
a parsimonious  parametrization of the probability functions  of the observed responses dictated by a RS. Several RSs
have been identified and studied \citep[e.g.,][]{baumgartner2001, van2013response} and here a model is introduced that enables capturing easily the most commonly encountered RS. In fact, in our approach, the observation probability functions, conditionally on the presence of a RS, depend on two parameters only, but  offer a great flexibility in the types of RSs that can be modelled such as
tendency to select at random categories (careless RS, CRS),
tendency to prefer positive response categories/answer with agreement (acquiescent RS, ARS),
or negative response categories (disacquiescent RS, DRS), middle/neutral categories (middle
RS, MRS), or extreme categories (extreme RS, ERS).
Other approaches to simultaneously tackle multiple RSs, for cross-sectional data,  rely on more complex models such as multi-trait models \cite[e.g.,][]{wetzel2015, falk2016} or item response models  \cite[e.g.,][]{ bockenholt2013, henninger2020, zhang2020validity} or latent class factor models \citep{kieruj2013response}.

Furthermore, novel is also in the use of stereotype logit models \citep{anderson1984regression} to investigate how covariates affect the initial and transition probabilities of the latent Markov chain. To our knowledge, such parsimonious models of sound interpretation have not
been previously used in the HMM framework.

In summary, our approach enables:
\begin{itemize}
	\item[(i)] the identification of groups of individuals  with a different dynamics of a latent categorical trait of interest taking into account the presence of RS driven responses,
	\item[(ii)] the accommodation of different RSs and their change over time,
	\item[(iii)] the use of a very parsimonious distribution of the responses affected by a RS,
	\item[(iv)]  the introduction of covariates affecting the initial/transition probabilities of both the latent construct and  the unobservable  response style indicator,
    \item[(v)] the use of stereotype logit models for the initial/transition probabilities of the latent construct.
\end{itemize}

The  proposed methodology  is   of interest in all longitudinal  surveys that model attitudes, opinions, perceptions or beliefs, that are indicators of non directly measurable and observable variables.  For example, in healthcare studies, patients are asked, at several occasions,  to give a subjective assessment of their health status or disability in daily living;
in marketing research,  customers are required to evaluate their satisfaction for services/products;
in socio-economic contexts,  citizens are invited to answer to what extent they agree or disagree
with sensitive topics (immigration, criminality, gender gap);
in environmental studies,  interviewees are asked to reveal their perception of the impact of climate changes and environmental risk.
In all these cases, the presence of RS effects cannot be ignored and
substantive  latent traits need to be measured taking into account   effects due to RSs.

To show the practical usefulness of our proposal, we investigate the evolution over time of the household financial capability   \citep[a broader term encompassing behaviour, knowledge, skills and attitudes of people with regard to managing their
financial resources, e.g.][]{zottel2013financial} as a latent psychological and behavioral trait  that influences the household's decision-making to face financial issues. The latent financial capability is here measured in terms of two observed indicators: the self-perceived \textit{ability} to  make ends meet and the self-report of perceived \textit{risk} related to financial investments. These  indicators
have great impact on the score measuring the financial capability, as defined according to the Organization for Economic Cooperation and Development
methodology \citep[survey][]{oecd}, applied by 36 countries and in Italy implemented by the Bank of Italy \citep{dfinancial}.

The structure  of the work is as follows. In Section~\ref{motiex}, the data of our motivating problem from a survey on financial conditions of  Italian households are introduced, and the issues to be tackled described.
In Section~\ref{RS}, the modeling of different response style effects in the longitudinal perspective through HMMs is proposed and the advantages of our approach are highlighted.
In Section~\ref{Sec.IntroMod},  latent  and  observation components of the proposed HMM are described in detail.  Alternative HMMs are examined in Section~\ref{alt}, most of them being special cases of the here presented model. Section~\ref{inf} is devoted to methodological contributions on: maximum likelihood estimators of the parameters, measures of goodness of fit and classification, and full-conditional residuals. In Section~\ref{sec:ex}, the
proposed model is fitted on the real data of Section~\ref{motiex}, implementing the developed estimation procedure and providing
answers to the questions raised in Section~\ref{motiex}. Concluding remarks are given in Section~\ref{concl}. Technical details on the methods to calculate the  standard errors are postponed in Appendix.

\section{SHIW Data-Bank of Italy}\label{motiex}

Our work meets the growing interest in the
households financial capability. The governments are now playing an active role in meeting the financial capability challenge. Initiative taking forward to increase capability are provided throughout starting from National Strategy for Financial Capability in UK \citep{britain2007financial}, to EU Commission \citep{valant2015improving}, National Financial Capability Study in USA \citep{lin2019state}, OECD \citep{oecd}, among others. European Commission recognised that individual financial and economic behavior is relevant to EU policy making process, and since 2009 incorporates behavioral insights into the design, implementation and
monitoring of EU policies \citep{van2013applying}. Psychological and behavioural aspects affecting people economic and financial decisions (studied as behavioral economics) are also inserted into practices to strengthen financial consumer protection \citep[as agreed in the action plan endorsed by G20/OECD,][]{lefevre2017behavioural}.

In this direction, we propose here to model the dynamics of the households' perception of their financial conditions, accounting for the way households disclose their perceptions, through HMM.
%In this direction, we investigate the evolution over time of the household financial capability as a latent psychological and behavioral trait  that influences the household's decision-making to face financial issues. The latent financial capability is here measured in terms of two observed indicators: the self-perceived \textit{ability} to  make ends meet and the self-report of perceived \textit{risk} related to financial investments. These  indicators
%have great impact on the score measuring the financial capability, as calculated according to the Organization for Economic Cooperation and Development
%methodology \citep[survey][]{oecd}, applied by 36 countries and in Italy implemented by the Bank of Italy \citep{dfinancial}.

The data are from the waves of the  Survey on Household Income and Wealth (SHIW). It is conducted by the Bank of Italy every two years since the 1960s to collect information about the income, wealth and saving of Italian households. Over the years, the survey has grown in scope and now it includes also aspects of households' economic and financial behaviour, furthermore since 2004 it contains information on attitude towards financial risk.
The data\footnote{All the data are available at https://www.bancaditalia.it/statistiche/tematiche/indagini-famiglie-imprese/bilanci-famiglie/distribuzione-microdati/index.html.} used refer to 1109 Italian households involved in all the waves from 2006 to 2016.

 We considered the items:
\begin{itemize}
	\item[--] $R_1$ reveals the perception of the household's financial \textit{ability} to make ends meet based on the answers of the head of the households to the question: \textit{Is your household's income sufficient to see you through to the end of the month....  very easily, easily,  fairly easily,  fairly difficultly,  with some difficulty,  very difficulty};
	\item[--] $R_2$ indicates the \textit{risk} perception in managing financial investments measured through the response to the question: \textit{in managing your financial investments, would you say you have a preference for investments that offer}: \textit{low returns, with no risk of losing the invested capital (risk averse);  a fair return, with a good degree of protection for the invested capital (risk tolerant), good-high returns, but with a fair-high risk of losing part of the capital (risk lover).}
\end{itemize}
\begin{figure}[h!]
	\begin{center}
		\includegraphics[width=1\linewidth]{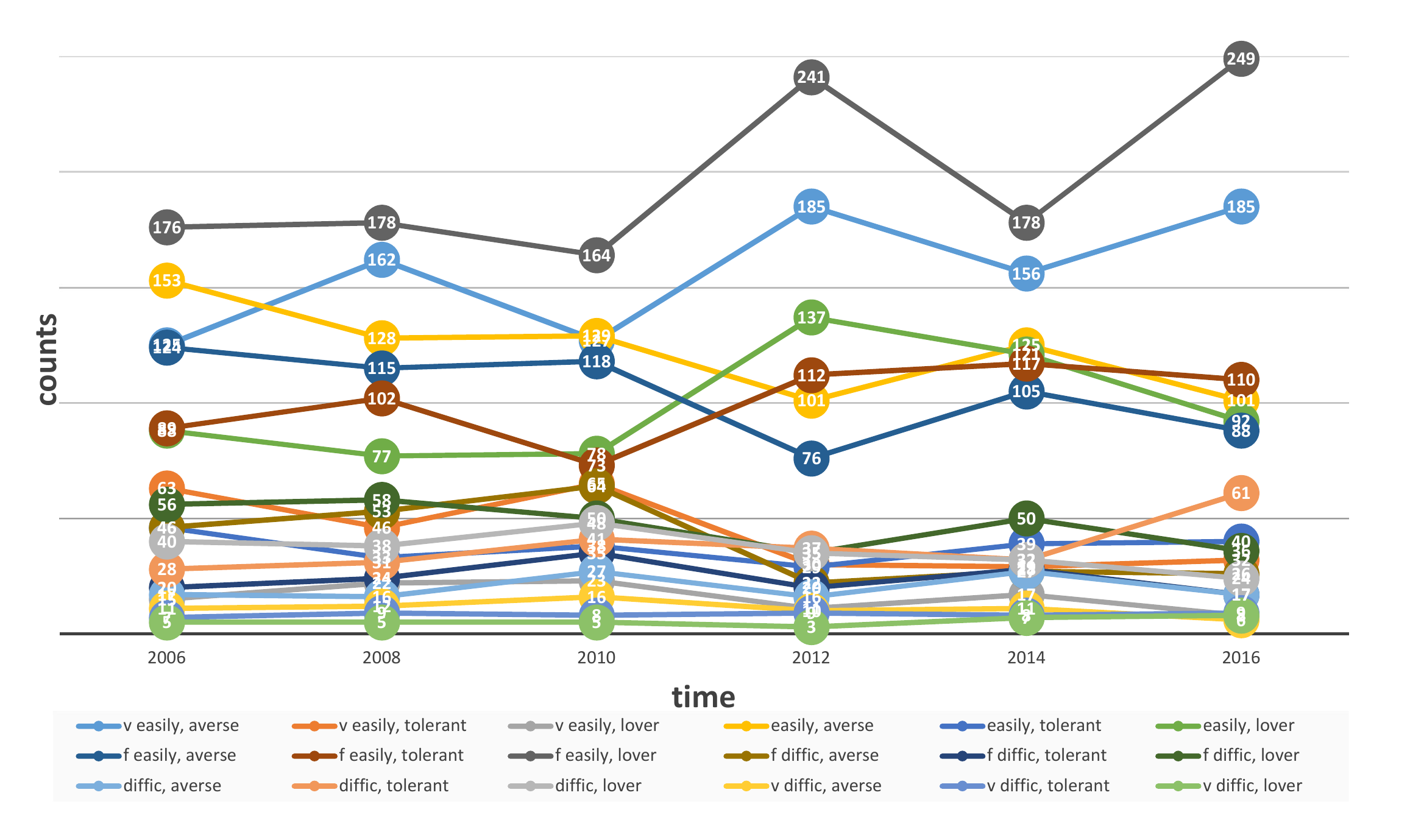}
		\caption{Frequencies of $(R_1,R_2)$ responses over time occasion}
		\label{Freq}
	\end{center}
\end{figure}
\hspace{-1.8 cm}\begin{figure}[h!]
	\begin{subfigure}{.52\textwidth}
		\centering
		% include first image
		\includegraphics[width=1.0\linewidth]{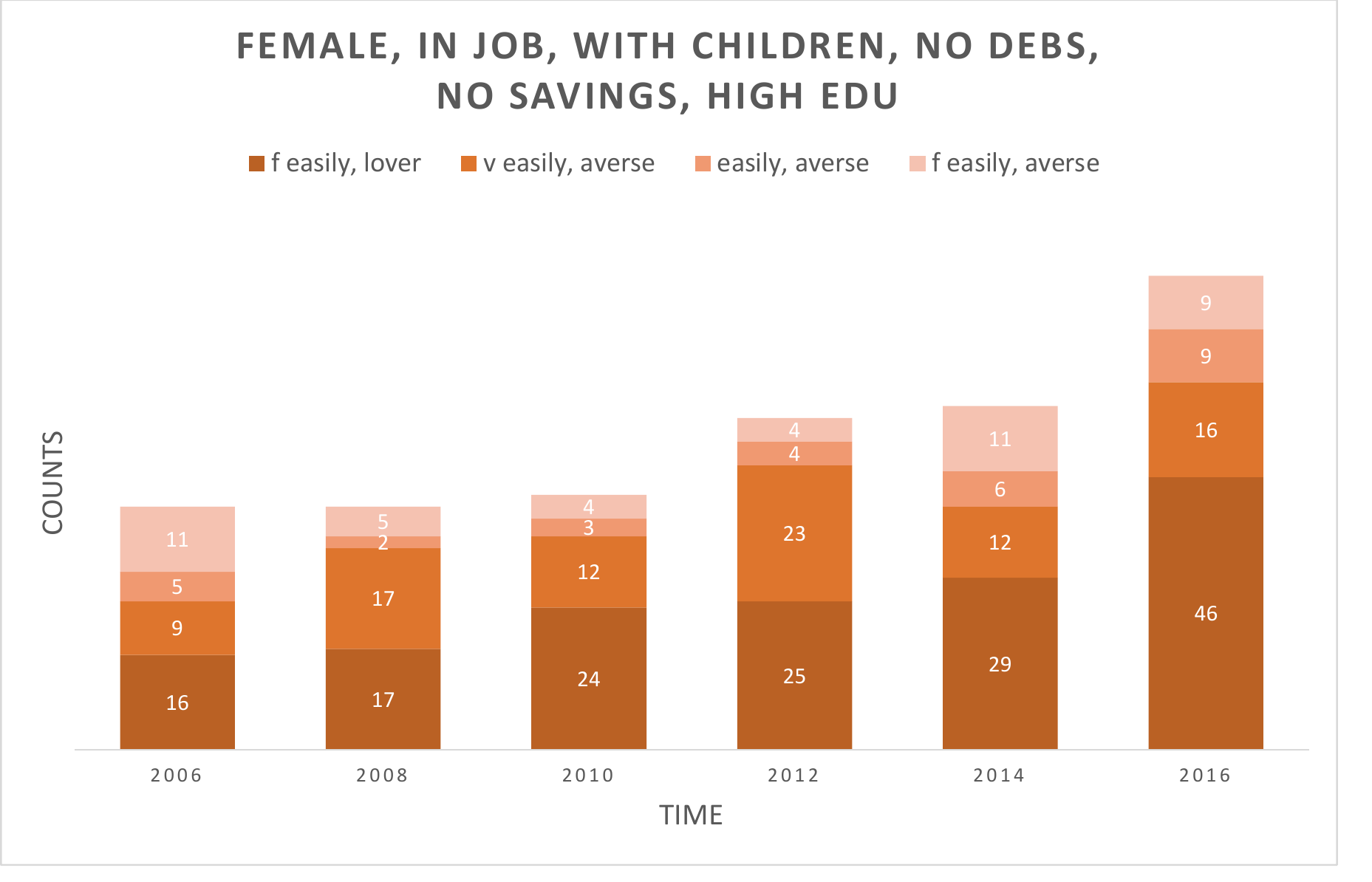}
		%\caption{Profile 1}
	\end{subfigure}
	\begin{subfigure}{.52\textwidth}
		\centering
		% include second image
		\includegraphics[width=1.0\linewidth]{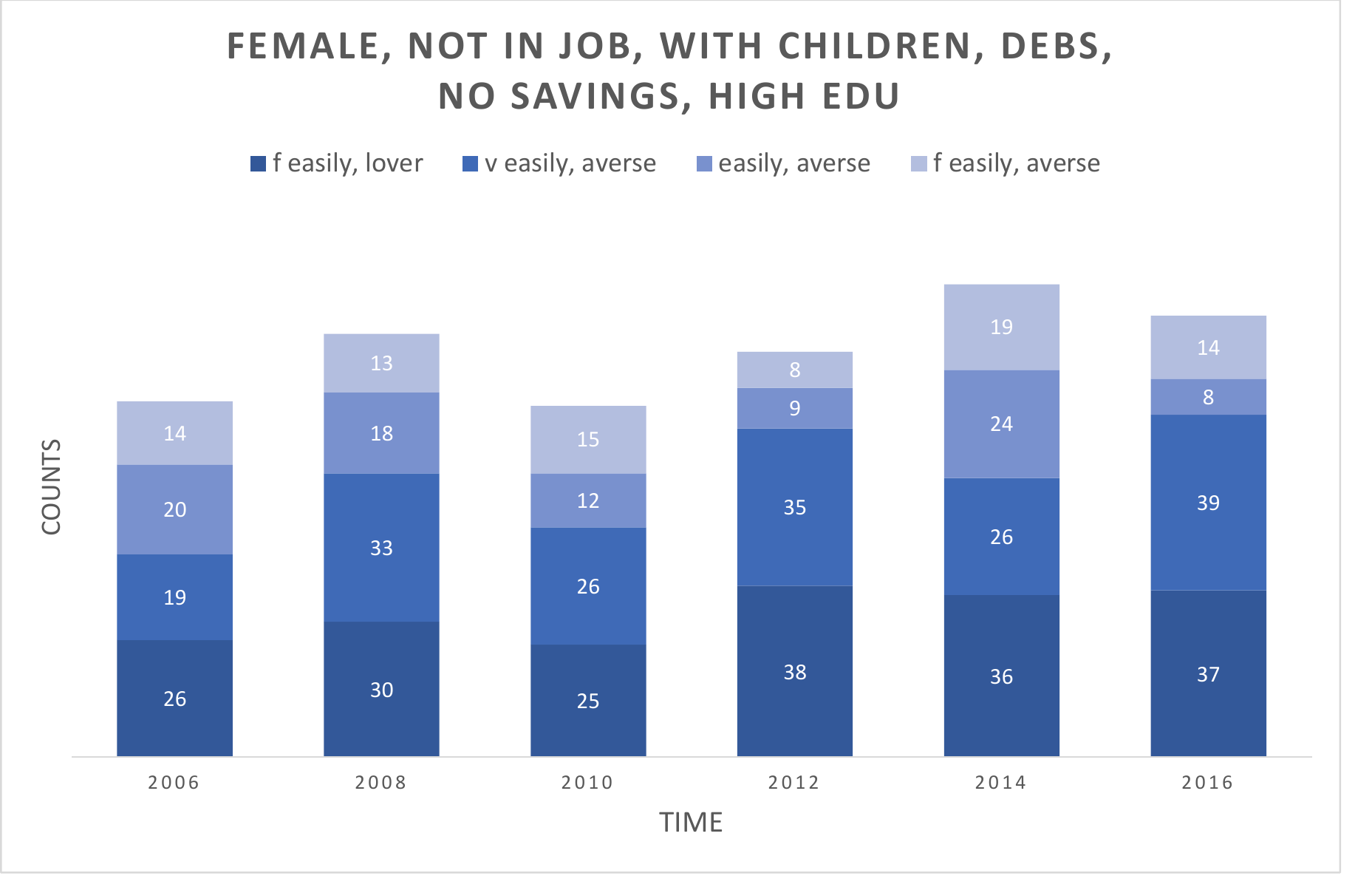}
		%\caption{Profile 2}
	\end{subfigure}
	\begin{subfigure}{.52\textwidth}
		\centering
		% include first image
		\includegraphics[width=1.0\linewidth]{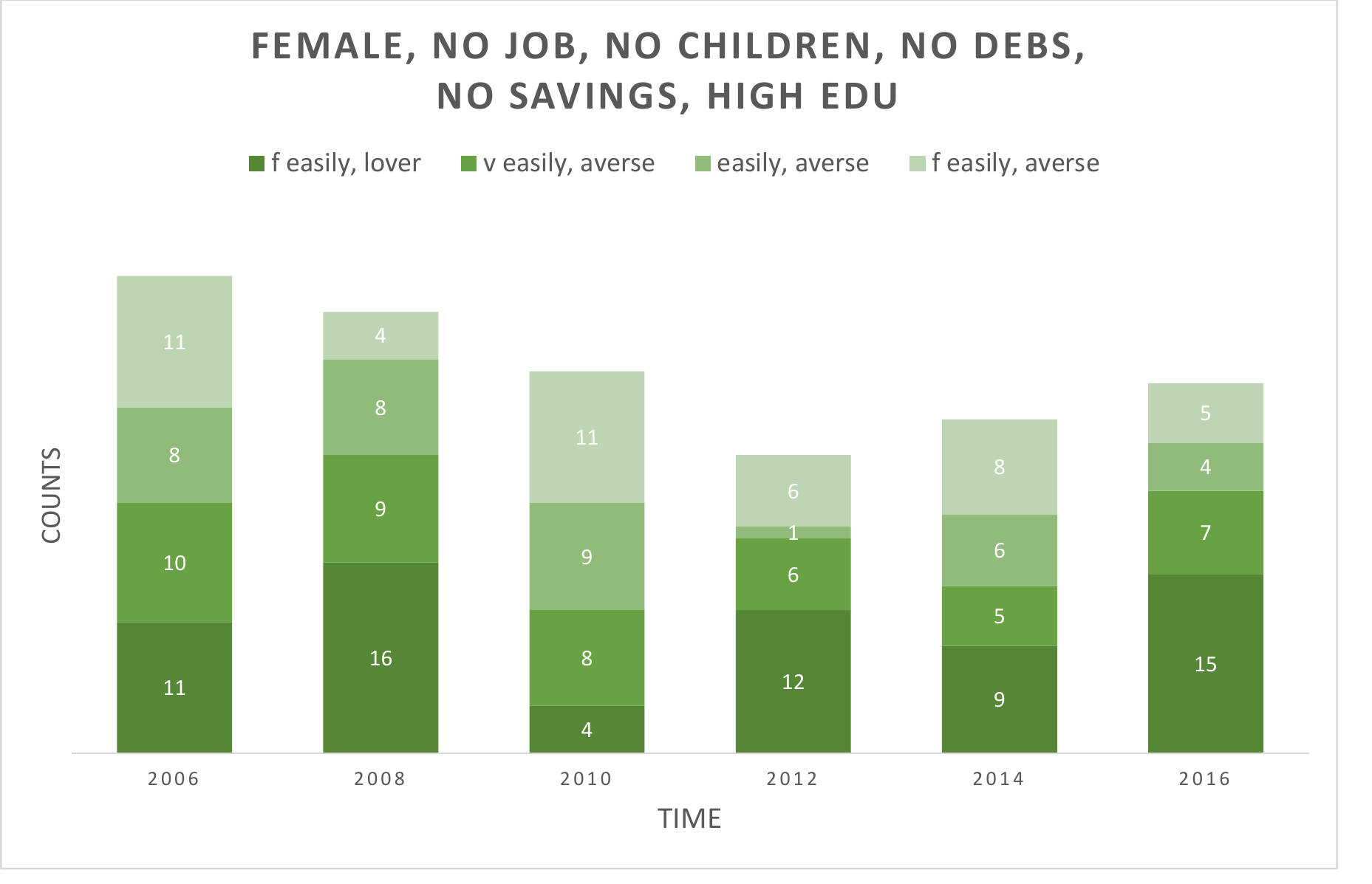}
		%	\caption{Profile 3}
	\end{subfigure}
	\begin{subfigure}{.52\textwidth}
		\centering
		% include second image
		\includegraphics[width=1.0\linewidth]{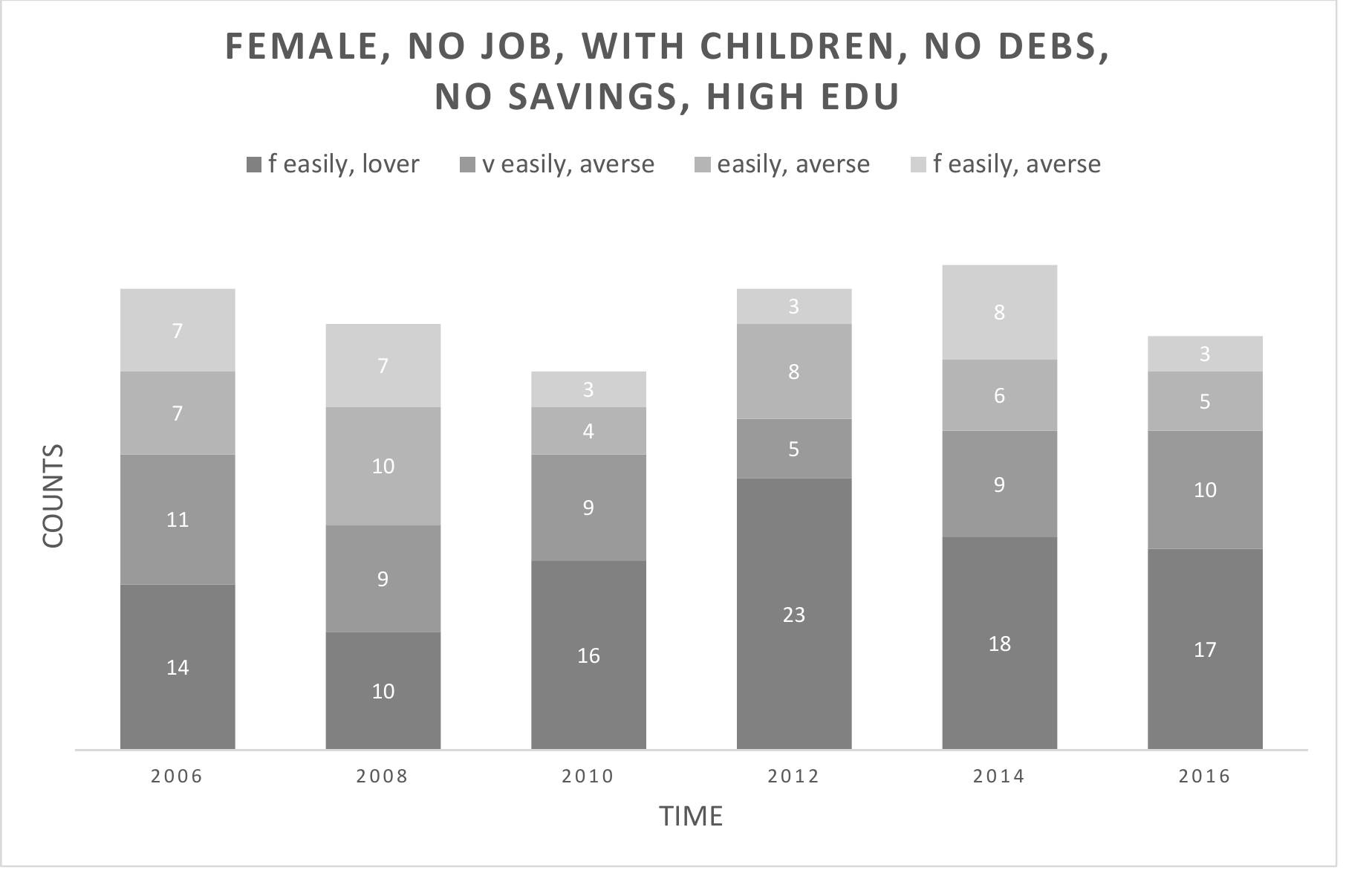}
		%	\caption{Profile 4}
	\end{subfigure}
\captionsetup{width=.85\textwidth}
	\caption{Frequencies of the most common responses over the six years for the four most present profiles of households}
	\label{profile}
\end{figure}

We focus on these two indicators, among others, since they strongly orient policy maker choices. In particular, insights into \textit{ability} helps to: developing effective programmes to educate people to manage their resources, reducing welfare dependency,   and identifying vulnerable groups of the
population for which targeted interventions can be
designed. The OECD, in the recent survey \citep{oecd}, recognises that large groups of citizens are lacking the necessary financial behavior and
financial resilience to deal effectively with everyday financial management. This is particularly
concerning at the time of the unfolding crisis as a result of the COVID-19 pandemic, which is likely
to put considerable economic and financial pressures on individuals and test their ability to
preserve their financial well-being. Moreover, to understand the financial capability is important to comprehend how households think and feel about the \textit{risk}s they face, \cite{slovic2010feeling}.   The risk perception is
an important determinant of protective behavior, as in general, the success of public
intervention programs is largely dependent on individual risk perception. Comprehension of the perceived risk may offer useful prompts for the design of effective investor education programmes and orient towards vulnerable individuals  preventive initiatives against bad financial decisions \citep[e.g.,][]{pidgeon1998risk, gentile2015financial, nguyen2019joint}.

\cite{jappelli}  investigated $R_1$ as an overall measure of financial distress and to describe from an economic point of view Italian households' saving and indebtedness behaviors.
Item $R_1$ has also been considered as a subjective component of a composite index for measuring the (latent) household financial condition in a time-invariant framework, used for describing features of household financial distress \citep{bialowolski2014index}. Furthermore, it is employed for measuring household financial capability by \citet{taylor2011measuring} and financial vulnerability by \cite{anderloni2012household}.
 \cite{de2021massive} used $R_2$ to stress how the risk preferences change over time and how the risk aversion has impact on the individual's occupational choice.

Some demographical and economical characteristics that can affect the degree of financial coping of a household are the covariates
gender (G): female ($27\%$), \textit{male} ($73\%$); job (J): self-employee (Jse, $10\%$), housekeeper/retired/student (Jhrs, $47\%$), \textit{employee} (Je, $43\%$)); children (CH):  with children ($34\%$), \textit{no children} ($66\%$); debts (D): with debts ($22\%$), \textit{no debts} ($78\%$); savings (S): with savings ($83\%$), \textit{no savings} ($17\%$); education (E): up to secondary school ($62\%$), \textit{over secondary school} ($38\%$), with the reference categories being in italics and the percentages referred to the initial year 2006.

The frequencies of all the 18 pairs of categories of the two items, $R_1$ and $R_2$, over the six years are represented in Figure~\ref{Freq}. The perceptions evidently change over time. The most commonly chosen responses are \textit{a fairly difficult} ability to make ends meet and a \textit{risk loving} behavior towards financial proposals. The choice falls frequently also on the pairs: \textit{very easily-averse, easily-averse, fairly easily-averse}. The number of households who selected these four most common responses is represented in Figure~\ref{profile}, over the years, for four groups of households, identified as those with the most representative profiles (most frequent configurations of the covariates among the 94 truly observed ones in the data at hand). These plots exemplify just a portion of the data on the three dimensions: responses $\times$ covariates $\times$ time.

In our context, responses $R_1$ and $R_2$ can be considered as the manifest expressions of the latent household's financial capability. Moreover, we believe that an unobservable answering behavior drives respondents that can reveal their perceptions in two ways: with awareness, when their answers reflect the respondents' true opinion, or according to a response style, when in doubt or reluctant to disclose their opinion they prefer extreme or middle categories of the rating scale, or according to their inclination they focus on the positive or negative side of the rating scale.

Our approach gives us various opportunities: (i) to describe simultaneously the dynamic behavior of respondents in the way of answering and in disclosing their perceived financial capability measured through the degree of difficulty or ease in matching monthly expenses with disposable income and their attitude toward risk of investments,  that change over time in line with \cite{schildberg2018risk} and \cite{de2021massive}; (ii) to investigate if the households feel and communicate their perceptions differently according to their demographic and socio-economic characteristics; (iii) to discriminate groups of respondents, with certain profiles, in the latent classes that identify various degrees of the latent financial capability, taking into account that they can answer with awareness or may prefer a response style.
Section~\ref{sec:ex} will shed light on these aspects of interest.

\section{Response Styles in Longitudinal Studies
}\label{RS}

RS mechanisms lead to biased measurement of the traits of interest that may influence seriously the results of a survey and thus be responsible for non-optimal decisions. Underlying RSs affect
all levels of the analysis of survey data, from being responsible for violations of the adopted model assumptions up to biased estimation of parameters and measures of interest, like correlations in cross-sectional survey data,
as shown, among others, by \cite{PIccSim} for CRS, \citet{dolnicar2009response} for ARS and ERS, \citet{tutzbergerEDU} for MRS and ERS.
Approaches aiming at better estimation of the original substantive trait,  by controlling for RSs,  are mostly based on mixtures models and employ latent variables   \citep[e.g.,][among many others]{grun2016response,  huang2016mixture, bockenholt2017response}. Several simulation studies provide evidence that ignoring response styles implies bias on the parameter estimates, see \cite{tutzbergerEDU, colombi2019hierarchical,colombi2021rating}, among others.

Though accounting for RSs in cross sectional studies has achieved considerably attention in the literature and effective models have been proposed, dealing with RSs in longitudinal data remains challenging. Questions on whether or to what extend RSs remain stable over time are still open.
This paper  investigates
whether RS behavior is an individual time invariant
feature  \citep{bachman1984yea, paulhus1991} or it is not necessarily consistent over time, depending on the measurement situation \citep{weijters2006response, aichholzer2013intra}.
In this regard,  the RS  is described through time-invariant and time-specific latent factors in \citep{weijters2010stability}.
A recent proposal in the direction of dynamic response styles is by \cite{soland2020response},  who concluded that the stability over time  of within-subject RS factors is not always justified, by  comparing multidimensional nominal response models
\citep{bolt2009applications}.

In the  context of longitudinal data, we tackle the problem of time dependence  of RSs within the latent variable context, considering two unobserved classes of responses:
aware (AWR) responses, i.e. not affected by any RS, and  RS driven responses,
assuming that an individual can switch over time from AWR to RS type of responses and vice versa.

Among  the  approaches  to modeling  longitudinal  categorical data \citep[representative sources are, for example,][]{molenberghs2005, hedeker2006, bergsma2009}, we resort to the family of hidden Markov  models for  their flexibility in modeling time dependence, based on sound assumptions, and their computational tractability. The novelty of the contribution is  the modeling of the temporal dynamic of rating  responses  with a bivariate latent Markov chain  that jointly models an unobservable construct of interest and an unobservable indicator of the respondent's form of answering (AWR or RS driven). The second latent component indeed allows us to describe how the RS behavior  dynamically changes over time, in contrast to other approaches
where the RS is thought as a continuous time-invariant latent trait \citep{billiet2008testing}.

\section{HMMs with two latent variables }\label{Sec.IntroMod}

Consider $r$  ordinal responses observed on $n$ units (subjects/items) at $T$ time occasions.
In particular, let $Y_{jit}$,  $Y_{jit}\in \mathcal C_j=\{1,\ldots, c_j\}$, denote the  $j$-th ordinal response variable, $j\in \mathcal R=\{1,\dots,r\}$, of the $i$-th unit, $i\in \mathcal I =\{1,\dots,n\}$, at the $t$-th occasion, $t\in \mathcal T =\{1,\dots,T\}$.
The responses are assumed to  reflect the levels of   unobservable  latent constructs
 $L_{it}$,  $i\in \mathcal I$, $t\in \mathcal T$, with finite discrete state space $\mathcal S_L=\{1,\dots,k\}$.
Furthermore, they can be observed under two latent regimes: awareness (AWR) and  response style (RS)  that are captured by binary latent variables $U_{it}$, $i\in \mathcal I$, $t\in \mathcal T$,  with state space $\mathcal S_U=\{1,2\}$, where 1 and 2 denote the RS and AWR states, respectively. For this, $U_{it}$ are called \textit{response style indicators}. The presence of the above mentioned two regimes is based on the idea that respondents either  manifest their true preference  or select  categories according to a RS (CRS, ERS, DRS, MRS, ERS).

 The proposal is a HMM defined by two components that describe the Markov chain  of the latent variables and the conditional distributions of the responses given the latent variables. The model will be referred to  as a HMM
with an RS component (RS-HMM). Next subsections are devoted to specifying the two  model components by parameterizing the observation probabilities and the initial/transition probabilities through suitable logit models. To avoid difficulties in  interpreting the results, covariates are assumed to affect only the distribution of the latent variables.  In our view, in fact, the covariate effect is  captured by the latent constructs $L_{it}$ which are indirectly observed trough the responses $Y_{jit}$.

\subsection{The latent model} \label{LAT}

The latent variables $L_{it}$ and $U_{it}$ are independent across units and, for every unit, the process $\{L_{it},U_{it}\}_{t\in\mathcal T}$ is assumed to evolve in time according to a first order bivariate Markov chain with states $(u,l)$, $u\in \mathcal S_U$, $l\in\mathcal S_L$.

For the sequel, let always $i\in\mathcal I$ and consider states $u, \ \bar{u} \in \mathcal S_U$ and $l, \ \bar{l} \in \mathcal S_L$.

\noindent The latent component of the  model is specified through its  initial and transition probabilities.
The initial probabilities ($t=1$) of the latent bivariate process $\{L_{it},U_{it}\}_{t\in\mathcal T}$  are
$\pi_{i1}(u,l)=P(L_{i1}=l,U_{i1}=u),$
and the transition probabilities are
$\pi_{it}(u,l|\bar u, \bar l)= P(L_{it}=l,U_{it}=u | L_{it-1}=\bar{l},U_{it-1}=\bar{u}),  t=2,\dots,T. $

 Furthermore,
\begin{equation}\label{margEMRS}
\pi^L_{it}(l|\bar u,\bar l)=P(L_{it}=l| L_{it-1}=\bar{l},U_{it-1}=\bar{u})
\end{equation}
denote the {\it marginal transition probabilities} for the latent variables $L_{it}$ and
\begin{equation}\label{condEMRS}
\pi_{it}^{U|L}(u|l,\bar u,\bar l) = \dfrac{\pi_{it}(u,l|\bar u, \bar l)}{\pi^L_{it}(l|\bar u,\bar l)}
\end{equation}
are the  transition probabilities of the latent RS indicators  $U_{it}$, conditioned on the transition $(\bar l,l)$  of the latent construct,
called for short as {\it conditional RS transition probabilities}.

The introduced  probabilities  are required to satisfy the following conditions:

\begin{itemize}
	\item[A1.] \textit{Granger non causality assumption}:
		$\pi^L_{it}(l|\bar u,\bar l)=\pi^L_{it}(l|\bar l),   t=2,\dots,T.$
	
	It states that $L_{it} \perp U_{it-1}|L_{it-1}$, i.e.\  the latent construct, given its past,  does  not depend on the past of the RS indicator.
	\item[A2.] \textit{Conditional independence of the current latent RS indicator  from the past of the latent construct:}
	$\pi_{it}^{U|L}(u|l,\bar u,\bar l) = \pi_{it}^{U|L}(u|l,\bar{u}),  t=2,\dots,T.$
	
	This restriction on the probabilities (\ref{condEMRS}) means that: $U_{it} \perp L_{it-1}|U_{it-1},L_{it}$, i.e.\ the current way of answering, depends on   its past and on the contemporaneous latent construct but not on the past of the latent construct.

	\item[A3.] \textit{Independence of the  latent processes at the initial time:}
	$\pi_{i1}(u,l)=\pi_{i1}^{U}(u)\pi^L_{i1}(l).$
\end{itemize}
	
Assumptions  A1 and A2 simplify the transition probabilities of the bivariate Markov chain $\{L_{it},U_{it}\}_{t\in\mathcal T}$ to
$\pi_{it}(u,l|\bar u, \bar l)=\pi^{U|L}_{it}(u|l,\bar{u})\pi^L_{it}(l|\bar l),  t=2,\dots,T,$ while A3 is used to reduce the number of parameters, but can be relaxed.

In the sequel, $\b x^{(m)}_i$ and $\b z^{(m)}_{it}$,  $m\in\{L,U\}$,  $t\in \{2,\ldots, T\}$, stand for the covariate row vectors influencing the initial and transition probabilities of the  latent variables for the $i$-th unit,  respectively.
The associated number of covariates is $p_1^{(m)}$ and $p_2^{(m)}$, respectively.
 Notice that the covariates for the transition probabilities can be time specific.

Under the  assumptions $A1-A3$, the  initial and transition probabilities of the  latent RS indicator and of the latent construct are specified by the following logit models:

\begin{itemize}
	\item[--] A linear baseline  logit model for the  initial probabilities of the latent construct:
	\begin{equation}\label{L1}
	\log \dfrac{\pi_{i1}^L(l)}{\pi_{i1}^L(1)}=\alpha_{0l}+\b \alpha_{1l}\tr \b x_{i}^{(L)}, \quad l=2,\dots,k.
	\end{equation}
  This model involves $(k-1)(1+p_1^{(L)})$ parameters.
	\item[--] A logit model for the initial probabilities of the RS indicator:
	\begin{equation}\label{L2}
	\log\frac{\pi^{U}_{i1}(2)}{\pi^{U}_{i1}(1)}=\bar\alpha_{0}+\bar{\b \alpha}_{1}\tr \b x_{i}^{(U)}.
	\end{equation}
	 This model has $(1+p_1^{(U)})$ parameters.
	\item[--]  A set of $|\mathcal S_L|=k$ linear baseline logit models for the marginal transition probabilities of the latent construct,  each having as reference category the state $\bar l$ of the previous occasion, i.e.\ for $\bar l\in \mathcal S_L$:
	\begin{equation}\label{L3}
	\log \dfrac{\pi_{it}^L(l|\bar l)}{\pi_{it}^L(\bar l|\bar l)}=\beta_{0l\bar l}+ \b \beta_{1l\bar l}\tr\b z_{it}^{(L)}, \quad l\in \mathcal S_L, \quad l\neq \bar l, \quad t=2,\dots,T.
	\end{equation}
	The total number of parameters for these models equals $k(k-1)(1+p_2^{(L)})$.
	\item[--]  A logit model for the conditional RS transition probabilities  for each possible RS state $\bar{u}$ of the previous occasion and for each current state $l$ of the latent construct:
	\begin{equation}\label{L4}
	\log\dfrac{\pi^{U|L}_{it}(2|l,\bar{u})}{\pi^{U|L}_{it}(1|l,\bar{u})}=\bar{\beta}_{0l \bar u}+ \bar{\b \beta}_{1l \bar u}\tr\b z_{it}^{(U)},\quad l\in \mathcal S_L, \quad \bar u\in \mathcal S_U, \quad t=2,\dots,T.
	\end{equation}
	The $2k$ models have in total $2k(1+p_2^{(U)})$ parameters.

\end{itemize}

 The number of parameters of models (\ref{L1}) and (\ref{L3}) for the latent construct is increasing in the number of states $k$, which is a draw back of these models.
More parsimonious models can be considered, alternative to (\ref{L1}) and (\ref{L3}), that provide sound interpretation options.
A convenient class of models for such purposes is that of stereotype logit models \citep{anderson1984regression, agresti2010analysis}, which we shall employ for modeling the initial and marginal transition probabilities of $L_{it}$.

The stereotype logit model for the initial probabilities, that can replace (\ref{L1}), is:
\begin{equation}\label{stereo_L1}
\log\frac{\pi^L_{i1}(l)}{\pi^L_{i1}(1)}=\alpha_{0l}+ \mu_l\b\alpha_{1}\tr \b x_{i}^{(L)}, \quad \mu_2=1, \quad l=2,\dots,k,
\end{equation}
where $\mu_l$, $l=2,\dots,k,$ are scores to be estimated. For identifiability purposes,  since the model is invariant under  scale transformations of the scores, we set $\mu_2=1$.
This model has $(k-1)+(k-2)+p_1^{(L)}$ parameters that is $(k-2)(p_1^{(L)}-1)$ parameters less than model (\ref{L1}).
 Model (\ref{stereo_L1}) imposes a special structure on the way the covariates affect the odds of any two categories of $L_{it}$. In particular, for any $l_1, l_2\in \mathcal S_L$, we have:
	\begin{equation}\label{stereo_L1_dif}
	\log\frac{\pi^L_{i1}(l_2)}{\pi^L_{i1}(l_1)} \ =\alpha_{0l_2}-\alpha_{0l_1} + (\mu_{l_2}-\mu_{l_1})\b\alpha_{1}\tr \b x_{i}^{(L)}, \quad \alpha_{01}=\mu_{1}=0,
	\end{equation}
	i.e. the effect of the covariates on the log-odds is proportional to the difference between the $\mu$-scores
	corresponding to the categories $l_1$ and $l_2$.

	The stereotype logit models for the transition probabilities, that can replace (\ref{L3}), are defined analogously as:
	\begin{equation}\label{stereo L3}
	\log \dfrac{\pi^L_{it}(l|\bar l)}{\pi^L_{it}(\bar l|\bar l)}=\beta_{0l\bar l}+\nu_{l\bar l} \b \beta_{1\bar l}\tr\b z_{it}^{(L)},
	\quad l\neq \bar l,\quad  l, \bar l\in\mathcal S_L,
	\end{equation}
	for $t=2,\dots,T$.  For $\bar l\neq 1$,   $\nu_{1\bar l}=1$, for  $\bar l=1$,   $\nu_{2\bar l}=1$ while the rest of the $\nu$-scores are parameters to be estimated.
	These models require $k(k-1+k-2+p_2^{(L)})$ parameters that is $k(k-2)(p_2^{(L)}-1)$ parameters less than  model (\ref{L3}).
	For  any $l_1\neq  l_2$, we have:
	\begin{equation}\label{stereo_eq}
\log \dfrac{\pi^L_{it}(l_1|\bar l)}{\pi^L_{it}( l_2|\bar l)} =\beta_{0l_1\bar l}-\beta_{0 l_2 \bar l} + (\nu_{l_1\bar l}-\nu_{l_2\bar l})\b \beta_{1\bar l}\tr\b z_{it}^{(L)},\quad \beta_{0 \bar l \bar l}=\nu_{\bar l\bar l}=0, \end{equation}
	i.e.\ the effect of the covariates on the log-odds is proportional to the difference between the $\mu$-scores  corresponding
	to the categories $l_1$ and $ l_2$.

If the   scores $\nu_{l\bar l}$ in (\ref{stereo L3}) are equal to 1, we obtain a more parsimonious model, according to which  the log odds (\ref{stereo_eq}) do not depend on covariates when $l_1 \neq l_2 \neq \bar l.$  According to this model  there is a covariate effect on the odds of a transition to a different state but this effect is the same for all the states different from the current one. Under this restriction,  (\ref{stereo L3}) simplifies to {\it parallel baseline logit models} for the  transition probabilities having $k(k-1+p_2^{(L)})$  parameters.

A different  simplification follows by assuming that  the  scores $\mu_l$, $\mu_1=0$,  $\mu_2=1$, $\nu_{l\bar l}$,  $\nu_{\bar l\bar l}=0$,   $\nu_{1\bar l}=1, \:\bar l\neq 1$,   $\nu_{2\bar l}=1, \:\bar l=1,$ are   linear functions of $l$, $l\in \mathcal S_L$. In this case, (\ref{stereo_L1}) and (\ref{stereo L3}) are equivalent  to {\it parallel adjacent categories logit models}  for the initial and transition probabilities having $(k-1)+p_1^{(L)}$ and $k(k-1+p_2^{(L)})$  parameters, respectively. Nevertheless, while the previous stereotype models are invariant with respect to permutations of the $k$ latent states, the parallel adjacent categories logit model is not and should be considered only in case the ordering of the latent classes is  known a priori.

Simplifying restrictions can also be introduced for the conditional RS
transition probabilities if, coherently with the idea that covariate effects are captured by the latent constructs $L_{it}$, the conditional RS transition probabilities are assumed time and subject invariant, that is:
\begin{equation} \label{hom}\pi_{it}^{U|L}(u|l,\bar{u})=\pi^{U|L}(u|l,\bar{u}), \quad i\in \mathcal I,  \quad  t=2,\dots,T.\end{equation}
%When (\ref{hom}) is satisfied, the conditional RS transition probabilities are called homogeneous,  otherwise heterogeneous.

\subsection{The observation model} \label{OBS}

 Let ${\bf Y}_i$ be the vector of the ordinal responses $Y_{jit}, j\in \mathcal R, \, t\in \mathcal T$  of unit $i$, $i\in \mathcal I$. Some independence assumptions specify the observation model:

\begin{itemize}
	\item[B1.] \textit{Subject independence}. The vectors ${\bf Y}_i$, $i\in \mathcal I$ are independent random vectors.
 \item[B2.] \textit{Hidden Markov assumption.}
	   For every unit  $i$ and occasion $t$, given  $\{L_{it},U_{it}\}_{t\in\mathcal T}$, the responses $Y_{jit}$, $j \in \mathcal R$, are independent from their own past and depend only on $(L_{it},U_{it})$.
	\item[B3.] \textit{Contemporaneous independence.}
	 For every unit  $i$, at any  occasion $t$, the responses $Y_{jit}$, $j \in \mathcal R$, are  independent given the current  state of the latent process $\{L_{it},U_{it}\}_{t\in\mathcal T}$.
	\item[B4.] \textit{Subject and time invariance}. The marginal probability functions of $Y_{jit}$, conditioned on the RS or AWR latent states $(u,l)$  are both time and subject invariant. That is, for $t\in \mathcal T$ and $i\in \mathcal I$,  it holds:
 $$f_{j|u}(y_j|l)=P(Y_{jit} = y_j | L_{it} = l, U_{it} = u), \quad j\in \mathcal R,
	\:u\in\mathcal S_U, \:l\in \mathcal S_L,  \:y_j\in \mathcal C_j. $$
\end{itemize}

\noindent Under the previous assumptions, the observation probability functions are parameterized by the following logit models (without covariates),  involving $k\sum_{j=1}^r (c_j - 1)+2rk$ parameters:

\begin{itemize}

	\item[--] Given the RS  regime,    every probability function  $f_{j|1}(y_j|l)$,  $j\in \mathcal R,$  $l\in\mathcal S_L$,
is specified by the   linear local logit model:
	\begin{equation}\label{O2}\log \frac{f_{j|1}(y_j+1|l)}{f_{j|1}(y_j|l)}=\phi_{0lj}+\phi_{1lj}s_j(y_j), \quad y_j=1,2,\dots,c_j-1.\end{equation}
	
		\item[--] Given  the  AWR regime of the  RS indicator,  every probability function  $f_{j|2}(y_j|l)$,  $j \in \mathcal R,$  $l\in\mathcal S_L$,  is parameterized by $c_j-1$  adjacent categories logits:
	\begin{equation}\label{O1}\log \frac{f_{j|2}(y_j+1|l)}{f_{j|2}(y_j|l)}=\varphi_{y_jl}, \quad y_j=1,2,\dots,c_j-1.\end{equation}
	%This model has $k\sum (c_j - 1)$ parameters.
\end{itemize}

The  $\phi_{0lj},\phi_{1lj} $ in (\ref{O2}) are parameters to estimate  and the scores   $s_j(y_j)$ are known constant defined as:
$s_j(y_j)=1$ for $y_j<c_j/2$, $s_j(y_j)=0$ for $y_j=c_j/2$, $s_j(y_j)=-1$ for $y_j>c_j/2$, $y_j = 1,2,\dots, c_j-1.$

These scores have been proposed by \cite{tutzbergerEDU} to extend the adjacent categories logit model to account for RS effects.

Parameter $\phi_{0lj}$ governs the skewness of the probability function  $f_{j|1}(y_j|l)$, so that it is symmetric with $\phi_{0lj}=0$, left and right skewed  with $\phi_{0lj}>0$ and $\phi_{0lj}<0$, respectively.

Increasing positive values of $\phi_{1lj}$ rise (decrease) the logits (\ref{O2}) for every $y_j$ which precedes (succeeds)  $\frac{c_j}{2}$.
	Hence,  for a fixed   $\phi_{0lj}$, greater positive values of $\phi_{1lj}$ make the response probability function $f_{j|1}(y_j|l)$, $y_j = 1,\dots,c_j,$ more concentrated around the  middle category $\frac{c_j+1}{2}$  (for $c_j$ odd) or the two middle categories $\frac{c_j}{2}, \frac{c_j}{2}+1$ (for $c_j$ even). With negative decreasing $\phi_{1lj}$, instead, the response probability function tends to be  more concentrated  on the extreme points. A formal definition of concentration around middle points of probability functions is given by \cite{colombi2021rating}.	
	
\begin{figure}[h!]
	\centering
	%\vspace{-1 cm}
	\includegraphics[width=0.7\linewidth]{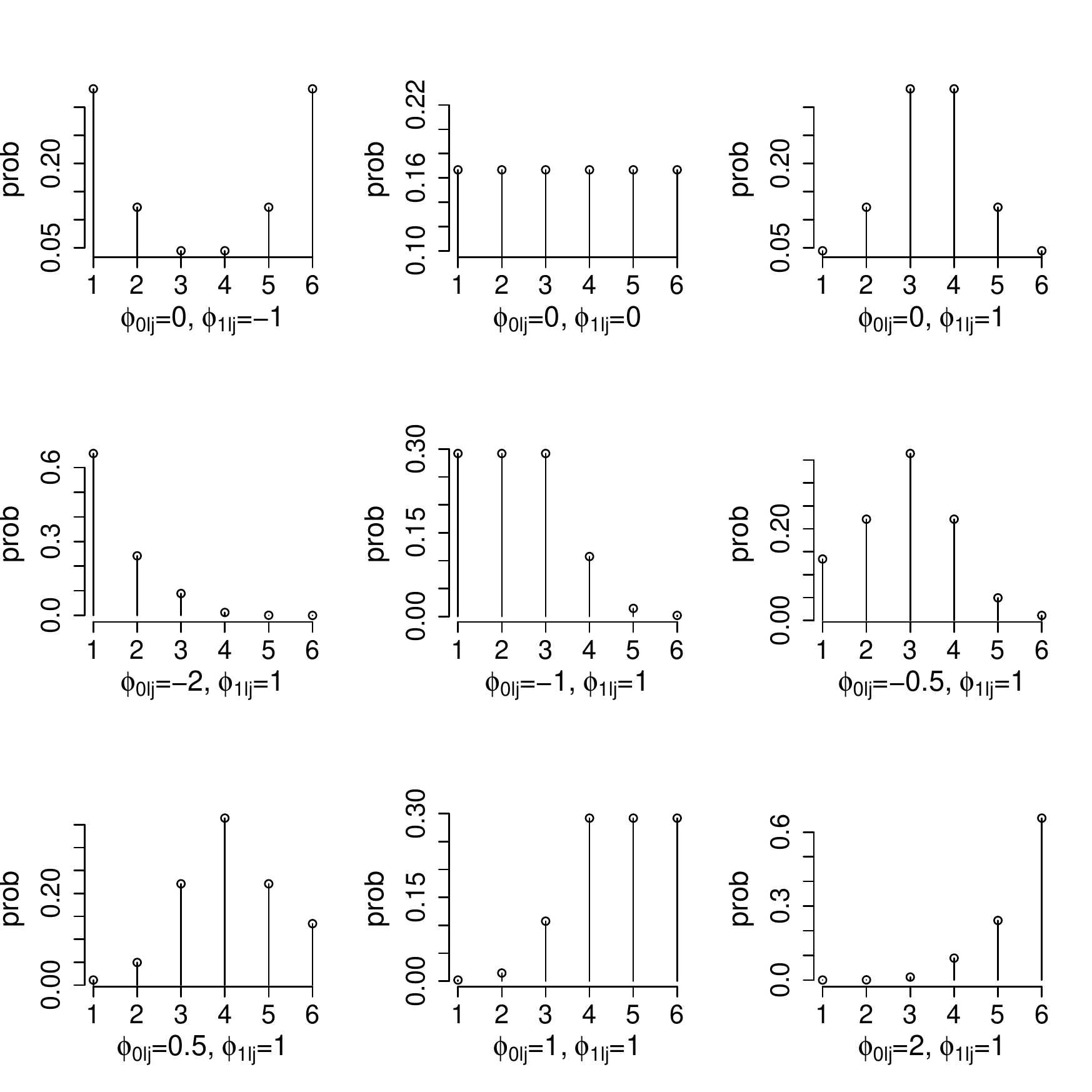}
	\captionsetup{width=.85\textwidth}
	\caption{Response probability functions  of respondents with ARS ($\phi_{0lj}=1,\phi_{1lj}=1$, or $\phi_{0lj}=2,\phi_{1lj}=1$),  DRS ($\phi_{0lj}=-2,\phi_{1lj}=1$, or $\phi_{0lj}=-1,\phi_{1lj}=1$), MRS ($\phi_{0lj}=0,\phi_{1lj}=1$ or $\phi_{0lj}=-0.5,\phi_{1lj}=1$ or $\phi_{0lj}=0.5,\phi_{1lj}=1$), ERS ($\phi_{0lj}=0,\phi_{1lj}=-1$) and CRS ($\phi_{0lj}=0,\phi_{1lj}=0$) patterns, for response categories ranging from $1 =$ \textit{strongly disagree} to $6 =$ \textit{strongly agree}}
	\label{Rplot9plots}
\end{figure}
	
The suitability of model (\ref{O2}) for describing ARS, DRS, MRS, ERS and CRS behaviors is justified by the fact that the  RS probability function defined by (\ref{O2}) can be unimodal only at the middle or extreme points categories of the response scale. In detail, for  $\phi_{1lj}>0$, the probability function  has a mode at the smallest category $y_j=1$  if  $\phi_{0lj}<-\phi_{1lj}$ (DRS) and at the highest category $y_j=c_j$ if  $\phi_{0lj}>\phi_{1lj}$ (ARS). For  $\phi_{1lj}>0$ and  $-\phi_{1lj}< \phi_{0lj}<\phi_{1lj}$, the mode is
at the middle (MRS) category $y_j=(c_j+1)	/2$ when $c_j$ is odd, while for even  $c_j$, the mode is at the middle   category $y_j=c_j/2$
(when $-\phi_{1lj}<\phi_{0lj}<0$) or at the middle category $y_j=c_j/2+1$ (when $\phi_{1lj}>\phi_{0lj}>0$).
If $\phi_{0lj}=-\phi_{1lj}$ ($\phi_{0lj}=\phi_{1lj}$), the previous modal categories and all the categories to the left (to the right) are equiprobable modes.

For  $\phi_{1lj}<0$, the probability function is U-shaped if  $\phi_{1lj}< \phi_{0lj}<-\phi_{1lj}$ (ERS) and  the mode corresponds to the smallest  (highest) category  when $\phi_{0lj}<0$ ($\phi_{0lj}>0$). If $\phi_{0lj}=0$, then the extreme categories are equiprobable modes. Finally, it is worth noting that $\phi_{1lj}=\phi_{0lj}=0$ gives the uniform distribution, commonly used to model CRS.

 Examples of the different shapes of the RS probability functions are illustrated in Figure~\ref{Rplot9plots}  to show the flexibility to model ARS, DRS, ERS, MRS, CRS.

\section{{Alternatives based on different  assumptions}} \label{alt}

By modifying assumption A2,  interesting models, proposed in the literature in different frameworks, can be obtained as special cases of RS-HMM.
These models deserve to be considered because they help us to understand the assumptions on the latent RS component of our approach.
Under the assumption $\pi_{it}^{U|L}(u|l,\bar u,\bar l) = \pi_{it}^{U}(u|\bar{u}), i \in \mathcal I, t\in\mathcal T$, more restrictive than A2, the Markov chains $L_{it},  t\in\mathcal T$,  and $U_{it}, t\in\mathcal T$ are independent for every $i \in \mathcal I$ (i.e.\ parallel Markov chains) and the RS-HMM model becomes a factorial HMM \citep{Zoubin1997} for longitudinal data. In this case, the RS latent component is unaffected by the latent construct component and vice versa, but both latent components influence the same observation component.

It is also  worth noting that under  the simplifying restriction $\pi^{U|L}_{it}(u|l,\bar u)$ $=\pi^{U|L}_{it}(u|l),$ of  memoryless  RS indicator (hereafter m.r.s.i),
	the RS-HMM model is equivalent to the HMM with the latent Markov variables $L_{it}, t\in\mathcal T$  only and with the univariate observation probability functions given by the mixtures:
	\begin{eqnarray}\label{memory_1}
		&f_{i1}(y_j|l)=\pi_{i1}^{U}(1)f_{j|1}(y_j|l)+\pi_{i1}^{U}(2)f_{j|2}(y_j|l), \quad t=1, \\
		&f_{it}(y_j|l)=\pi_{it}^{U|L}(1|l)f_{j|1}(y_j|l)+\pi_{it}^{U|L}(2|l)f_{j|2}(y_j|l), \quad t>1, \label{memory_t}
	\end{eqnarray}
	for $l\in \mathcal S_L$, $y_j\in \mathcal C_j$, $j \in \mathcal R$ where $f_{j|1}$ and $f_{j|2}$ are the marginal probability functions of $Y_{jit}$, conditioned on the RS or  AWR latent states, respectively, which are time and unit invariant (s.\ B4 in Section \ref{OBS}).
HMMs with mixtures in the observation components have been considered by \cite{volant2014hidden} in a general framework where the mixtures can have a different number of components depending on the state of the latent construct.

Under the stronger restriction $\pi^{U|L}_{it}(u|l,\bar u)=\pi^{U|L}_{it}(u)$, the RS indicators $U_{it}$, $ t\in\mathcal T$ are a sequence of independent random variables for every unit $i \in \mathcal I$.
This restriction has been used by \cite{DeSantis2011} to model \emph{zero inflation} in longitudinal count data.
 According to their approach, %of this authors,
the latent variables  $U_{it}$ %are used to
indicate presence or lack of a \emph{structural zero} while the $L_{it}$ are associated to different  rates or intensities  of Poisson counts.

Under the even stronger  restriction $\pi_{it}^{U}(u|\bar{u})=d_{\bar{u}}(u), i \in \mathcal I, t\in\mathcal T$,  $d_{\bar{u}}(\bar{u})=1$, $d_{\bar{u}}({u})=0,$ if $u \neq \bar{u}$, the RS component becomes a time invariant random component that impacts all the repeated observations on a subject.
This is a case of a HMM with a discrete random effect on the observation component, according to the terminology by \cite{bartolucci2012latent}, \cite[see also][]{Maruotti}.

Compared to the previous  sub-models,  our model is more general and flexible
since it does not assume time invariance for the RS component and, more generally, independence of the RS component from its own past or the latent construct.

Furthermore, a  model, not nested in the RS-HMM,  is obtained if A2 is replaced by the Granger non causality condition
$\pi_{it}^{U|L}(u|\bar u,\bar l) = \pi_{it}^{U}(u|\bar{u}), t=2,\dots,T,$
which is analogous to the Granger non causality assumption A1. This model, according to which  each latent  variable does not Granger cause the other one, is a special case of the graphical multiple HMMs introduced by \cite{ColGio2015}. The drawback of this model is that, under the  two non Granger causality conditions, the transition probabilities $\pi_{it}(u,l|\bar u, \bar l)$ do not have a closed expression and must be computed numerically as a function of the probabilities $\pi_{it}^{U}(u|\bar{u})$, $\pi^L_{it}(l|\bar l)$ and a set of $k-1$ odds ratios defined on the bivariate transition probabilities. See \cite{ColGio2015} for more details on these Granger non causality conditions and on  a marginal parametrization that can be used in this context.

\section{Inference} \label{inf}
Let $\b \theta$ denote the vector of all the parameters of the latent and observation models.
For example, in the simple case of a memoryless model with $k=2$, no covariates and one response with four categories, it is:
$\b \theta= (\alpha_{01},\bar{\alpha_0},\beta_{21},\beta_{12},\bar{\beta}_{01},\bar{\beta}_{02},\varphi_{11},\varphi_{21},\varphi_{31},
\varphi_{12},\varphi_{22},$ $\varphi_{32},\phi_{01},\phi_{11},\phi_{02},\phi_{12}).$
Hereafter, procedures to provide maximum likelihood estimates (MLE) of these parameters and standard errors are illustrated.

\subsection{Estimation via an EM algorithm} \label{EM}

The latent binary   variable $d_{it}^{(1)} (u,l)$ is equal to 1 when the $i$-th unit (subject) is at time $t$ in state $(u,l)$ and the latent binary  variable  $d_{it}^{(2)} (u,l;\bar u ,\bar l)$ is 1 if at time $t$, $t>1$, the $i$-th subject  is in state $(u,l)$ while at occasion $t-1$  was in $(\bar u, \bar l)$, $l, \bar l \in \mathcal S_L$, $u, \bar u \in \mathcal S_U$.
Moreover, the observable binary variable $d_{jit}(y_j)$ is equal to 1 if at time $t$  the category $y_j$ of $Y_{jit}$, $j\in \mathcal R$, is observed on the $i$-th individual, $i\in \mathcal I$.

If the above binary latent variables were observable, the parameters could be estimated by maximizing the following \emph{complete log-likelihood} (i.e. the joint log-likelihood of the observations and the latent variables):
\begin{eqnarray}\label{complete}\nonumber
&\ell^*(\b \theta)=\sum_{i=1}^{n}\sum_{l=1}^{k}\left[\sum_{u=1}^{2}d_{i1}^{(1)}(u,l)\right]\log\pi^L_{i1}(l)+ \\ \nonumber
&\sum_{i=1}^{n}\sum_{u=1}^{2}\left[ \sum_{l=1}^{k}d_{i1}^{(1)}(u,l) \right]\log\pi_{i1}^{U}(u)+\\
&\sum_{\bar l=1}^{k}\left\{\sum_{i=1}^{n}\sum_{t=2}^{T}\sum_{ l=1}^{k}\left[\sum_{u=1}^{2}\sum_{\bar u=1}^{2}d_{it}^{(2)} (u,l;\bar u ,\bar l)) \right ]\log \pi^L_{it}(l|\bar l)\right\}+\\ \nonumber
&\sum_{\bar u=1}^{2}\sum_{l=1}^{k}\left\{\sum_{i=1}^{n}\sum_{t=2}^{T}\sum_{u=1}^{2}\left[\sum_{\bar l=1}^{k}d_{it}^{(2)}(u,l;\bar u ,\bar l)) \right ]\log\pi_{it}^{U|L}(u|\bar{u},l)\right\}+\\ \nonumber
&\sum_{j=1}^{r}\sum_{l=1}^{k}\left\{\sum_{y_j=1}^{c_j}\left[\sum_{i=1}^{n}\sum_{t=1}^{T}d_{it}^{(1)}(1,l)d_{jit}(y_j)\right]\log f_{j|1}(y_j|l)\right\} +\\ \nonumber
&\sum_{j=1}^{r}\sum_{l=1}^{k}\left\{\sum_{y_j=1}^{c_j}\left[\sum_{i=1}^{n}\sum_{t=1}^{T}d_{it}^{(1)}(2,l)d_{jit}(y_j)\right]\log f_{j|2}(y_j|l)\right\}
,
\end{eqnarray}
where $f_{j|1}$ and $f_{j|2}$ are provided in (\ref{O2}) and (\ref{O1}).

As the latent variables are not observable and it is not  easy to maximize the marginal log-likelihood, obtained by summing the joint log-likelihood  over all the possible realizations of the latent indicators, it is common, in the context of HMMs, to use the EM algorithm, to compute the maximum likelihood estimates. Details on the EM algorithm in the context of HMMs are presented in many papers and books. See \cite{Bart2015} for a presentation specific to  the context of longitudinal data.
Every iteration of the EM algorithm is composed by two steps: the  Expectation (E) step and the Maximization (M) step.
With respect to our model, in the E step the following expected values are computed:
\begin{equation}\label{delta}\delta_{it}^{(1)} (u,l;\bar{\b \theta})=E_{obs}(d_{it}^{(1)} (u,l)), \quad \delta_{it}^{(2)} (u,l;\bar u ,\bar l;\bar{\b \theta})=E_{obs}(d_{it}^{(2)} (u,l;\bar u ,\bar l))),\end{equation}
where $E_{obs}()$ is the  expected value taken conditionally on the observed values of the responses $Y_{jit}$ and on the covariates and given
the current value $\bar{\b \theta}$ of the parameters.
The previous expected values are computed  by the Baum-Welch forward-backward algorithm \citep[Ch.\ 4]{Zucchini2009}.

In the M step, the following conditional expectation of the complete log-likelihood function is maximized in order to obtain an updated $\bar{\b \theta}$:
\begin{eqnarray}\nonumber
&Q(\b \theta|\bar{\b \theta})=E_{obs}(\ell^*(\b \theta))=\sum_{i=1}^{n}\sum_{l=1}^{k}\left[\sum_{u=1}^{2}\delta_{i1}^{(1)}(u,l;\bar{\b \theta})\right]\log\pi^L_{i1}(l)+ \label{l1}\\ \nonumber
&\sum_{i=1}^{n}\sum_{u=1}^{2}\left[ \sum_{l=1}^{k}\delta_{i1}^{(1)}(u,l;\bar{\theta}) \right]\log\pi_{i1}^{U}(u)+\label{l2}\\ \label{EM:Q}
&\sum_{\bar l=1}^{k}\left\{\sum_{i=1}^{n}\sum_{t=2}^{T}\sum_{ l=1}^{k}\left[\sum_{u=1}^{2}\sum_{\bar u=1}^{2}\delta_{it}^{(2)} (u,l;\bar u ,\bar l);\bar{\b \theta}) \right ]\log \pi^L_{it}(l|\bar l)\right\}+\label{l3}\\ \nonumber
&\sum_{\bar u=1}^{2}\sum_{l=1}^{k}\left\{\sum_{i=1}^{n}\sum_{t=2}^{T}\sum_{u=1}^{2}\left[\sum_{\bar l=1}^{k}\delta_{it}^{(2)}(u,l;\bar u ,\bar l);\bar{\b \theta}) \right ]\log\pi_{it}^{U|L}(u|\bar{u},l)\right\}+\label{l4}\\ \nonumber
&\sum_{j=1}^{r}\sum_{l=1}^{k}\left\{\sum_{y_j=1}^{c_j}\left[\sum_{i=1}^{n}\sum_{t=1}^{T}\delta_{it}^{(1)}(1,l;\bar{\b \theta})d_{jit}(y_j)\right]\log f_{j|1}(y_j|l)\right\}+\\ \nonumber
&\sum_{j=1}^{r}\sum_{l=1}^{k}\left\{\sum_{y_j=1}^{c_j}\left[\sum_{i=1}^{n}\sum_{t=1}^{T}\delta_{it}^{(1)}(2,l;\bar{\b \theta})d_{jit}(y_j)\right]\log f_{j|2}(y_j|l)\right\}.
\end{eqnarray}

Note that $Q(\b \theta|\bar{\b \theta})$ is obtained from the complete log-likelihood by replacing $d_{it}^{(1)} (u,l)$ and $d_{it}^{(2)} (u,l;\bar u ,\bar l)$ with their expected values (\ref{delta}).

The six addends of (\ref{EM:Q}), corresponding to the models specified by (\ref{L1}) or (\ref{stereo_L1}), (\ref{L2}), (\ref{L3}) or (\ref{stereo L3}), (\ref{L4}) and (\ref{O1}-\ref{O2}) of Sections~\ref{LAT} and \ref{OBS}, depend on disjoint  subsets of the vector $\b \theta$  and can be maximized separately. The maximization of the sixth addend is simple as there is a closed form for the maxima.
Moreover, the first addend is equivalent to the ML estimation of the logit model (\ref{L1}) or its stereotype variant (\ref{stereo_L1}), and the third and fifth terms simplify to the estimation  of  $k(r+1)$ separate logit models described by (\ref{L3}) or (\ref{stereo L3}) and (\ref{O2}). A similar remark applies to the second and fourth addends and the logit models defined by (\ref{L2}) and (\ref{L4}), respectively.
The terms within curled brackets correspond to the log-likelihoods of  the  logit models that can be maximized separately.   In the first two addends, the curled brackets are omitted as only one logit model is involved. The expected values within squared brackets play the role of observed frequencies.

If the model is correctly specified, the estimates  of the standard errors can be based either on the matrix of second derivatives of the log-likelihood function (observed information matrix, in short OIM), see \cite{Bart2015}, or on the outer products of the individual contributions to the score functions  \citep[outer product information matrix, OPIM, or BHHH estimate,][]{berndt1974estimation}.
When the model is misspecified, the information matrix equivalence does not hold and the standard errors have to be calculated using the so called Sandwich matrix \citep{white1982}, say SDW. Alternatively, standard errors can be computed using the boostrap (BOOT) technique. Technical details are given in Appendix.

All the R functions, for the estimates and standard errors (with the four mentioned methods) are available from the authors.

	\subsection{Goodness of fit and classification} \label{sec:GOF}
	
	The goodness of fit testing and model selection in latent Markov models for longitudinal data is not straightforward, since standard asymptotic results for test statistics may not hold.
	The use of Akaike's information criterion (AIC) or Bayesian information criterion (BIC) is a broadly used and accepted procedure. In particular, for HMM, the use of BIC dominates, even though
	its theoretical properties are not clear \citep[e.g.,][]{Bart2009, Zucchini2009}.
	
	Furthermore, there have been proposed in the literature normalized  indices for assessing the overall fit of a model. For example, \cite{Bart2009} used the index:
	\begin{equation}\label{R2}
	R^2=1-\exp\left\{2[\hat\ell_0 - \ell(\hat{\b\theta})]/(n \cdot r)\right\},
	\end{equation}
  for assessing the fit of the model against the independence  model characterized by  $k=1$  and no RS effects,	with  $\sum_{j=1}^r (c_j-1)$ parameters and  log-likelihood function $\hat\ell_0$. It holds $R^2 \in [0, 1]$,  with higher values indicating a better fit.
	
Indices can be introduced  for measuring the quality of classification and the distinguishability of the latent classes   as well; \cite{Bart2009} proposed an index based on the posterior probabilities of the latent classes, which in our set-up is:	\begin{equation}\label{S}
		S_k = \frac{\sum_{i=1}^n\sum_{t=1}^T (\delta^*_{it}-1/2k)}{(1-1/2k)nT},
		\end{equation}
		with $\delta^*_{it}$ being, for unit $i$ at time $t$, the maximum with respect to $(u,l)$ of the posterior latent class probabilities $\delta_{it}^{(1)} (u,l;\bar{\b \theta})$, introduced in (\ref{delta}).
		Measure	 $S_k$ lies between 0 and 1, where 1 represents certainty in classification and  a perfect separation among latent classes, while values close to 0 indicate that most of $\delta^*_{it}$ are close to $1/2k$, that is like choosing the classes at random.
		This index is very suitable for our context where the observed responses are manifest realizations of the latent variables, therefore a good quality in terms of separation of  the $2k$ latent states is crucial.
		
		In line with the literature which ignores the answering behavior, we can measure the quality of the separation of the latent construct states marginally with respect to $U$, so that (\ref{S}) reduces to:
	$$S^{L}_k = \frac{\sum_{i=1}^n\sum_{t=1}^T (\delta^{L}_{it}-1/k)}{(1-1/k)nT}, \quad with \quad \delta^L_{it}=\max\limits_{l \in \mathcal S_L} \sum_{u \in \mathcal S_U}\delta_{it}^{(1)} (u,l;\bar{\b \theta}).$$

		Moreover, in our context,  the distinguishability among the $k$ states of the latent construct can be interestingly measured at the AWR and RS regimes separately. The $S_k$ index is specified for this aim  as follows:
		$$S^{L|RS}_k = \frac{\sum_{i=1}^n\sum_{t=1}^T (\delta^{L|RS}_{it}-1/k)}{(1-1/k)nT}, \quad with \quad  \delta^{L|RS}_{it}=\max\limits_{l \in \mathcal S_L} \frac{\delta_{it}^{(1)} (1,l;\bar{\b \theta})}{\sum_{l^* \in \mathcal S_L}\delta_{it}^{(1)}
		(1,l^*;\bar{\b \theta})},$$
		
		$$S^{L|AWR}_k = \frac{\sum_{i=1}^n\sum_{t=1}^T (\delta^{L|AWR}_{it}-1/k)}{(1-1/k)nT},\quad with \quad  \delta^{L|AWR}_{it}=\max\limits_{l \in \mathcal S_L} \frac{\delta_{it}^{(1)} (2,l;\bar{\b \theta})}{\sum_{l^* \in \mathcal S_L}\delta_{it}^{(1)}
		(2,l^*;\bar{\b \theta})}.$$

It can also be of  interest  to measure the ability to discriminate between AWR and RS behaviors, regardless of the latent construct. For this aim, the measure (\ref{S}) is modified as:
$$S^{U}_k = \frac{\sum_{i=1}^n\sum_{t=1}^T (\delta^U_{it}-1/2)}{(1-1/2)nT},\quad with \quad \delta^U_{it}=\max\limits_{u \in \mathcal S_U} \sum_{l \in \mathcal S_L}\delta_{it}^{(1)} (u,l;\bar{\b \theta}).$$
Finally, the concern can be directed to measure how well separated are the two responding regimes, in every class of the latent construct. An insight in this sense is given by:
		$$S^{U|L = l}_k = \frac{\sum_{i=1}^n\sum_{t=1}^T (\delta^{U|L}_{itl}-1/2)}{(1-1/2)nT}, \quad with \quad  \delta^{U|L}_{itl}=\max\limits_{u \in \mathcal S_U} \frac{\delta_{it}^{(1)} (u,l;\bar{\b \theta})}{\sum_{u^* \in \mathcal S_u}\delta_{it}^{(1)}
		(u^*,l;\bar{\b \theta})}, \quad l \in \mathcal S_L.$$

\subsection{Residual analysis} \label{res}

After the selection of a reasonable model according to indices of goodness of classification, and indices for judging the overall fit of the model,  a residual analysis which detects features of the data not captured by the model has to be carried out.

We assess the adequacy of the selected  model by analysing \textit{full-conditional}  residuals, introduced in the context of HMMs by \cite{Buckby2020},  as \emph{exvisive} residuals. Full-conditional residuals are  an alternative to the \textit{forecast} or \textit{predictive} residuals \citep{Buckby2020}.
The difference is that in full-conditional residuals, the expected values of observed counts at time $t$ are taken given all the other observations while in forecast residuals they are taken given the observations before time $t$.
Full-conditional residuals are more useful in evaluating goodness of fit while forecast residuals are more helpful  to assess the predictive accuracy of the model.
In the application that follows, we use \textit{Pearson full-conditional residuals}, whose technical details are given below.

To simplify the notation, let $\b x_{it}= (\b x^{(U)'}_i,\b z_{it}^{(U)'},\b x^{(L)'}_i,\b z_{it}^{(L)'})\tr$ be
the set of covariates for individual $i$ at time $t$, $i \in \mathcal I$, $t\in \mathcal T$.
Let $\mathcal D_t=\{\b x_1,\b x_2,\dots,\b x_{d_t}\}$ be the set of different configurations of covariates observed at time $t\in \mathcal T$ and  $\mathcal D=\cup_t \mathcal D_t$.
Moreover  $\mathcal C$ is the set of the $c=\prod_j c_j$ different configurations of the  responses.
For every vector $\b y_{it},\: \b y_{it} \in \mathcal C$,  of the r responses of unit $i$ at time $t$, we define \emph{the rest of $\b y_{it}$ }as  $\mathcal Y_{it}^-=\{\b y_{i 1},\b y_{i 2}\dots,\b y_{i t-1},\b y_{i t+1},\dots,\b y_{iT}\}.$
For  $\b y \in \mathcal C,$  $i\in\mathcal I$, $t\in\mathcal T$, the indicator
$d_{it}(\b y)=1$ if $\b y_{it} = \b y, d_{it}(\b y)=0$ otherwise, is defined
and summing over units the counts
$n_{t}(\b y,\b x)=\sum_{i:\b x_{it}= \b x}d_{it}(\b y)$ are obtained  for every $\b x \in \mathcal D_t$.

We  introduce a residual for every  $\b x \in \mathcal D_t$, $t\in\mathcal T$, and $\b y \in \mathcal C$ by comparing the previous counts with their expected values defined below.

Let $f_{it}(\b y|\mathcal D,\mathcal Y_{it}^-)$, $\b y \in \mathcal C$ , $i \in \mathcal I,$ $t\in \mathcal T$  be the joint probability density function (pdf) of the responses given the covariates and the rest of $\b y_{it}$. The computation of this pdf is described by \cite{Buckby2020}, by \cite{Zucchini2009} in the  related  context of pseudo residuals, and can be obtained as a by product of the Baum-Welch algorithm.
Starting from these pdf, we define the following conditional expected values of the counts $n_{t}(\b y,\b x)$:
$$\mu_t(\b y,\b x)=\sum_{i:\b x_{it}= \b x}f_{it}(\b y|D,\mathcal Y_{it}^-),$$ for every $\b x \in \mathcal D_t$,  $t\in\mathcal T$, $\b y \in \mathcal C$.

Accordingly, the following full-conditional Pearson residuals are introduced:
\begin{equation}\label{eq-Pearson-res}
\rho_t(\b y,\b x)=\frac{n_{t}(\b y,\b x)-\mu_t(\b y,\b x)}{\sqrt{\mu_t(\b y,\b x)}} ,
\end{equation}
for every $\b x \in D_t$, $t\in \mathcal T$, $\b y \in \mathcal C$.
Plotting full-conditional Pearson residuals is an useful   tool to investigate the lack of fit of the model and to highlight particular features of the data. Standardizing these residuals is  possible in theory but the computation of the standard errors is  not an easy analytical and computational task. This could  be done by the methods  used in  \cite{Tit2009} for HMMs in continuous time but, an  in depth-study is needed to asses the feasibility in presence of many residuals.

For every time occasion $t$, $t\in \mathcal T$, and every observed covariate configuration $\b x$, the squared full-conditional Pearson residuals sum to the corresponding Pearson's chi-squared statistic
%\begin{equation} \label{chires}
$\chi^2_t(\b x)=\sum_{\b y \in \mathcal C}\rho_t(\b y,\b x)^2.$
In this paper, the averages of full-conditional Pearson residuals over the $c$ response configurations, i.e.\ $\frac{\chi^2_t(\b x)}{c}, \b x \in \mathcal D_t, t\in\mathcal T,$ are used to summarize the comparison of the estimated cell probabilities under the assumed model with the observed proportions.

In applications of multivariate responses, practical interest may lie on univariate responses  $Y_j$ or bivariate responses $(Y_j,Y_{j'})$,
with $j\neq j'$, $j, j' \in \mathcal R$. In such cases, residuals (\ref{eq-Pearson-res}) can be marginalized to:
\begin{equation*}
\rho_t^+(\utilde{\b y},\b x)=\frac{\sum_{j \in \mathcal R \setminus \utilde{\mathcal R}} n_{t}(\b y,\b x)-
\sum_{j \in \mathcal R \setminus \utilde{\mathcal R}} \mu_t(\b y,\b x)}
{\sqrt{\sum_{j \in \mathcal R \setminus \utilde{\mathcal R}} \mu_t(\b y,\b x)}} ,
\end{equation*}
where $\utilde{\b y}$ is a configuration  of the responses of interest and $\utilde{\mathcal R}$ the associated set of indices. The consideration of marginalized residuals is also useful in case of sparsity in response configurations.

\section{SHIW data analysis} \label{sec:ex}

We applied the proposed models to the panel data from the Survey on Household Income and Wealth described in Section~\ref{motiex} to answer the questions raised there. The  household's financial capability  (or  condition) is the latent trait of interest measured trough the ability to make ends meet $R_1$ and the perceived financial risk $R_2$, with covariates gender (G), job (J), children (CH), debts (D), savings (S), education (E), cf Section~\ref{motiex}.

\subsection{Model selection}

\begin{table}[htbp]
	\centering
	\caption{The maximum value of the log-likelihood function ($loglike$),  the number of states $k$, the number of parameters,  BIC,
	$S_k$ and $R^2$ values are reported for  models defined by different hypotheses on the transition probabilities.  }
	\resizebox*{1\textwidth}{!}{\renewcommand\arraystretch{1.2}
		\begin{tabular}{rrccccccc}
			\cmidrule{2-3}          & \multicolumn{2}{c}{\textbf{Hypotheses on  transition probabilities }} &       &    &   &       & \multicolumn{1}{r}{}  \\
			\midrule
			\multicolumn{1}{l}{{Model}} & \multicolumn{1}{c}{$\pi^L_{it}(l|\bar l)$} & {$\pi_{it}^{U|L}(u|l,\bar{u})$} & {k} & $loglike$ & {n.\ par.} & {BIC} & {$S_k$} & ${R^2}$\\
			\midrule
			\multicolumn{1}{l}{M1} & \multicolumn{1}{c}{unrestricted logit models} & \multicolumn{1}{c}{heterogeneous, m.r.s.i.} & 2    & -15119.64 & 70 & 30730.07 & 0.707 & 0.805\\
			&       &       & 3      & -14716.95& 129 & 30338.34 & 0.699 & 0.864\\
			&       &       & 4      & -14497.80& 204& 30425.89 & 0.692 & 0.888\\
			&       &       & 5      & -14309.60 & 295 & 30687.50 & 0.713 & 0.906\\
			\midrule
			\multicolumn{1}{l}{M2} & \multicolumn{1}{c}{unrestricted logit models} & heterogeneous & 2      &-14773.11 &86 &30149.18 & 0.798 & 0.857\\
			&       &       & 3      & -14447.88 & 153 &29968.47 & 0.760 & 0893\\
			&       &       & 4      & -14289.60& 236 & 30233.85 & 0.720 & 0.903\\
			&       &       & 5      & -14191.18 & 335 & 30731.12 & 0.715 & 0.915\\
			\midrule
			\multicolumn{1}{l}{M3} & \multicolumn{1}{c}{stereotype logit models} & heterogeneous & 2     & -    &   -   &  - & - & -\\
			&       &       & 3      &-14465.32& 129& 29835.08 & 0.764 & 0.892\\
			&       &       & 4      & -14330.35& 176&29894.68 & 0.717 & 0.904\\
			&       &       & 5      &-14282.73& 227& 30157.01 & 0.681 & 0.908\\
			\midrule
			\multicolumn{1}{l}{M4} & \multicolumn{1}{c}{parallel baseline logit models} &  heterogeneous & 2      &-14773.11 &86  &30149.18 & 0.798 & 0.857\\
			&       &       & 3      &-14466.80& 126& 29817.02 & 0.768 & 0.891\\
			&       &       & 4      & -14350.83& 168& 29879.55 & 0.716 & 0.902\\
			&       &       & 5      & -14307.23& 212& 30100.84 & 0.691 & 0.906\\
			\midrule
			\midrule
			\multicolumn{1}{l}{M5} & \multicolumn{1}{c}{unrestricted logit models} & \multicolumn{1}{c}{homogeneous, m.r.s.i.} & 2     & -15340.33& 49& 31024.21 & 0.695 & 0.762 \\
			&       &       & 3      & -14860.61& 101& 30429.35& 0.690 &0.845\\
			&       &       & 4      & -14625.49& 169& 30435.88 & 0.645 & 0.875\\
			&       &       & 5     & -14457.99& 253& 30689.81 & 0.660 & 0.892\\
			\midrule
			\multicolumn{1}{l}{M6} & \multicolumn{1}{c}{unrestricted logit models} & homogeneous & 2      & -14833.29& 58& 30073.23& 0.800 & 0.849\\
			&       &       & 3      & -14512.86& 111& 29803.96&0.759 & 0.887\\
			&       &       & 4      &-14364.24& 180& 29990.50& 0.702 & 0.901\\
			&       &       & 5      &-14265.3& 265& 30388.58& 0.691 & 0.909\\
			\midrule
			\multicolumn{1}{l}{M7} & \multicolumn{1}{c}{stereotype logit models} & homogeneous & 2     & -     & -     & - & -& -\\
			&       &       & 3      &-14532.10& 87& 29674.19 & 0.760& 0.885\\
			&       &       & 4      & -14413.66& 120& 29668.66 & 0.695 & 0.897\\
			&       &       & 5      & -14354.00& 157& 29808.76 & 0.661 & 0.902\\
			\midrule
			\multicolumn{1}{l}{M8} & \multicolumn{1}{c} { parallel baseline logit models}  &   homogeneous & 2      & -14833.29& 58& 30073.23 & 0.800 & 0.849\\
			&    &      & 3  & -14534.76& 84& 29658.46 & 0.764& 0.885\\
			&       &      & 4 & -14427.96& 112& 29641.18 & 0.713 & 0.895\\
			&       &       & 5      & -14370.93& 142& 29737.46& 0.697& 0.900\\
			\midrule
	\end{tabular}}
	\label{compmod}
	
\end{table}%

Models based on different hypotheses on the latent transition probabilities are compared in Table~\ref{compmod}, each one considered for an increasing number of latent states.
 When the initial or transition probabilities  depend on the covariates are said to be heterogeneous  otherwise they are  homogeneous.

In the models of Table~\ref{compmod}, the initial probabilities  $\pi^{L}_{i1}(l)$ are modelled through stereotype logits (models M3, M4, M7, M8) when a stereotype model or a parallel baseline logit model is used for the transition probabilities of the latent construct, otherwise they are modelled by  unrestricted logit models
(models M1, M2, M5, M6). RS initial probabilities are always assumed to depend on the covariates to capture heterogeneity  in the answering behavior at the beginning.

The minimum BIC corresponds to the model M8 with $k = 4$ states defined by stereotype models (\ref{stereo_L1}) for the latent construct initial probabilities and parallel baseline logit models (\ref{stereo L3}) with scores $\nu_{l\bar l}=1$, $l \neq \bar l$, for the transition probabilities, and no covariate effects on the RS conditional transition probabilities, specified in  (\ref{L4}).
Model M8 with $k=3$ is the second best according to the BIC criterion. Both the models M8 with $k=3$ and $k=4$ have a very high value of $R^2$, so they fit similarly and well enough the data at hand, stressing that the dependence of the responses on time and covariates is supported by the data. Nevertheless, there is evidence of quite overfitting for Model M8 ($k=4$), since the conditional response probabilities in two states are not easily distinguishable. Looking at the goodness of the classification of units into the latent classes, measured by the $S_k$ index (\ref{S}), it results that model M8 with $k=3$ has $S_3=0.757$ greater than $S_4=0.712$ obtained for $k=4$, thus the simple model seems to better separate the latent classes.
Moreover, the results of all the variants of the $S_k$ index, i.e.\ measures $S^L_k$, $S^U_k$, $S^{L|RS}_k$, $S^{L|AWR}_k$, $S^{U|L=l}_k$ with $l \in \mathcal{S_L}$ (Section~\ref{sec:GOF}), illustrated in Table~\ref{tab:Sindices}, confirm the superiority of M8 with $k=3$ over the analogous model with $k=4$ in terms of distinguishing the states of the latent financial capability within the two groups of AWR and RS respondents, and also the greater ability to distinguish the AWR and RS behaviors, marginally and conditionally on the latent classes $l$, except for $l=1$  only.

\begin{table}[h]
	\centering
	\captionsetup{width=.85\textwidth}
	\caption{Results of indices of quality of classification illustrated in Section~\ref{sec:GOF} for models M8 with $k=3$ and $k=4$ states of the latent construct}
	\resizebox*{0.8\textwidth}{!}{\renewcommand\arraystretch{1.2}
		\begin{tabular}{cccccccccc}
			\cmidrule{1-10}
			k     & $S_k$   & $S^L_k$  & $S^U_k$ & $S^{L|RS}_k$  & $S^{L|AWR}_k$ & $S^{U|l=1}_k$& $S^{U|l=2}_k$ & $S^{U|l=3}_k$ & $S^{U|l=4}_k$ \\\cmidrule{1-10}
			3     & 0.757 & 0.824 & 0.747 & 0.847 & 0.837  & 0.823 & 0.813& 0.821&\\\cmidrule{1-9}
			4     & 0.712 & 0.783 & 0.738 & 0.801 & 0.805  & 0.872& 0.736& 0.785& 0.823\\\cmidrule{1-10}
	\end{tabular}}
	\label{tab:Sindices}%
\end{table}%

Therefore, the latent construct - the households financial capability - is reasonably chosen with three states meaning that households can be grouped according to whether they feel financially confident ($l=1$), financially fair ($l=2$), financially distressed ($l=3$).
The choice of model M8 implies that the transition probabilities of the latent construct are well described under the parsimonious model, where the effects of covariates on the transition probabilities depend on the previous latent state $\bar{l}$ but do not change over the current latent state $l$. Other less restrictive hypotheses do not fit better, also when combined with a smaller number of latent states.

 The chosen model is then compared with the  latent Markov models, with three states and six states, proposed by \cite{bartolucci2012latent}, say B, that are alternative to our model but do not account for a distinct latent variable representing the answering behavior. The comparison with the B model with $k= 3$ ($loglike = 14883.72$, $npar=85$, $BIC=30363.39$) validates the idea that an underlying binary latent variable that distinguishes AWR and RS respondents is coherent with the data at hand, thereby strengthening empirically the usefulness of our approach. Moreover, the comparison with the alternative B model with $2k=6$ ($loglike =-14612.20$, $npar=322$, $BIC=31482.02$) confirms that the restrictions hypothesized in our model on the transition probabilities and on the RS probability functions are reasonable for the analyzed data.

\begin{figure}[h!]
	\begin{center}
		\includegraphics[width=0.9\linewidth]{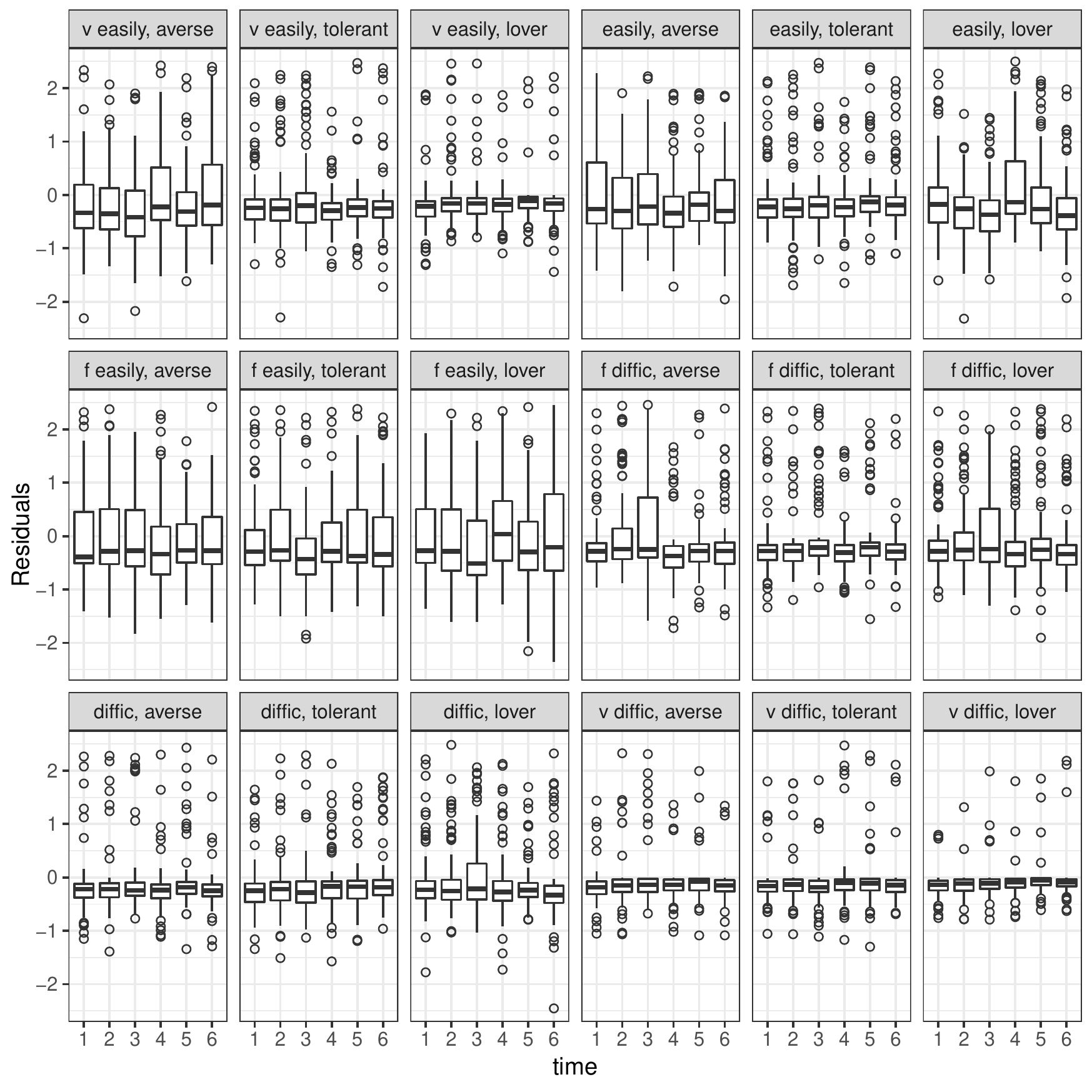}
	\end{center}
	\caption{Box plots of residuals for occasion time and configurations of responses }
	\label{res36}
\end{figure}

To complete  the assessment of the chosen model we carry out a residual analysis.
Figure~\ref{res36} illustrates the $3 \times 6 \times 6$ box plots of the full-conditional Pearson residuals (hereafter residuals), described in (\ref{eq-Pearson-res}), calculated within the 6 time occasions for every combination of the categories of the two responses. Box plots, within each time occasion and responses configuration, correspond to different covariate profiles.

All the residuals, across all time occassions, have  very small values around zero with $median = -0.210$, $Q_1= -0.437$, $Q_3=-0.025$, $mean = -0.011$, $sd = 0.93$. In particular, $95.6\%$  of them are between -2 and 2.
Overall, 11$\%$ are out of whiskers in the box plots of Figure~\ref{res36} while only 22 residuals in total are greater than 5 ($2.5$ \textperthousand).
The maximum residual corresponds to  the profile of a male, with a job (self-employee or employee), no children, no debts, no savings and a low educational level.
%Rplotres3x6white.pdf
A slightly larger dispersion appears for residuals (s. Fig~\ref{res36})
corresponding to the choice \textit{fairly easily} for $R_1$ combined with all possible responses on risk perception $R_2$ (\textit{averse}, \textit{tolerant} and \textit{lover}).

\begin{figure}[h!]
	\begin{center}
		\hspace{-2 cm}	\includegraphics[width=0.5\linewidth]{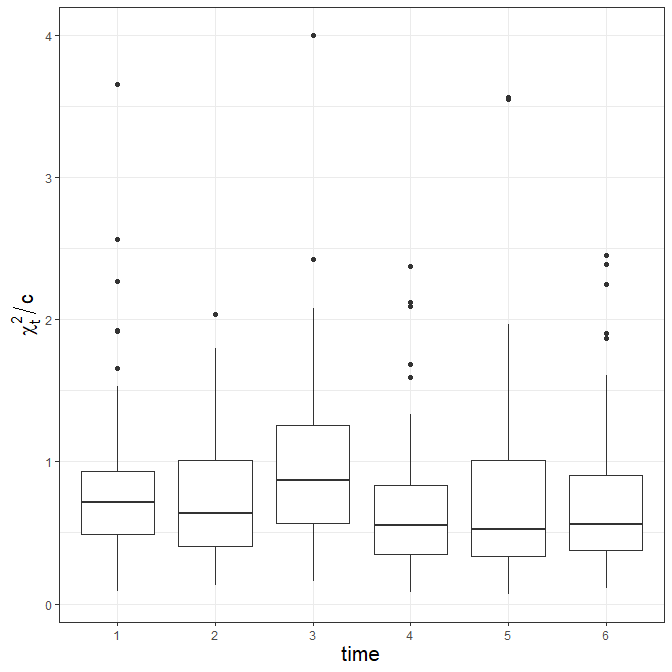}
	\end{center}
\captionsetup{width=.85\textwidth}
	\caption{Box plots of the averages of full-conditional Pearson's residuals for every covariate configuration at every time occasion}
	\label{chi_TC}
\end{figure}

Furthermore, we calculated the averages of the  full-conditional Pearson's residuals (Section~\ref{res});
Figure~\ref{chi_TC} illustrates the box plots of these averages, for every covariate configuration at every time occasion.
We observe quite small values overall, with the exception of two points having average of
Pearson's residuals greater or equal to 4, both at time $t=3$ (one of them not shown on the plot for better visualization purposes).
They correspond both to female respondents with quite opposite profiles.
In one of them they are  not  self-employee, with no children-debs-savings and a low education, while in the other,
they have a job (self-employee or employee), children-debs-savings and a high educational level.

\begin{table}[h]
	\centering
	\captionsetup{width=.85\textwidth}
	\caption{Estimates (standard errors) of parameters $\phi_{0lj}$ and $\phi_{1lj}$ of the RS probability functions in the stata $l=1,2,3$ and responses $R_j$, $j=1,2$ }
	\resizebox*{0.90\textwidth}{!}{\renewcommand\arraystretch{1.2}
		\begin{tabular}{clccc}
			\toprule    \multicolumn{1}{r}{} &       & \multicolumn{1}{l}{financially confident $(l=1)$} & \multicolumn{1}{l}{financially fair $(l=2)$} & \multicolumn{1}{l}{financially  distressed $(l=3)$} \\
			\midrule
			\multirow{2}[1]{*}{$R_1$} & $\phi_{0l1}$ & -1.4642 (0.1106) & 1.2743 (0.1446) & 3.8061 (0.0897) \\
			& $\phi_{1l1}$ &  2.5575 (0.1256) & 2.2705 (0.1431) & 2.9025 (0.1132) \\
			\cmidrule{2-5}    \multirow{2}[2]{*}{$R_2$} & $\phi_{0l2}$ & 1.4101 (0.1052) & -1.5093 (0.9448) & -0.9618 (0.0539) \\
			& $\phi_{1l2}$ & 0.4533 (0.1611) & -0.4743 (0.9496) & -0.1526 (0.0824) \\
			\bottomrule
		\end{tabular}%
		\label{tablephi}}%
\end{table}%

\begin{figure}
	\centering
	\vspace{-1 cm}\includegraphics[scale=0.5]{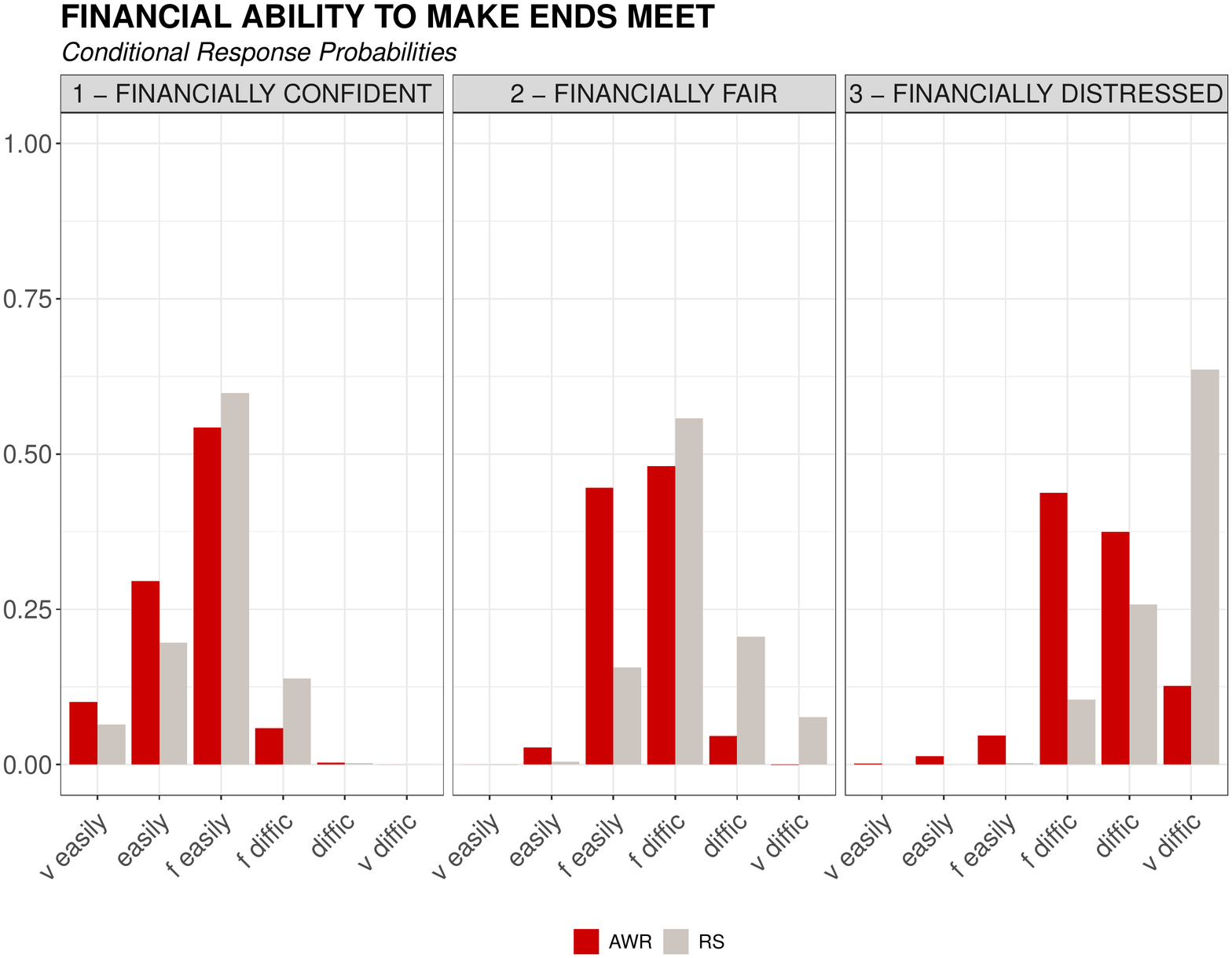}
	\includegraphics[scale=0.5]{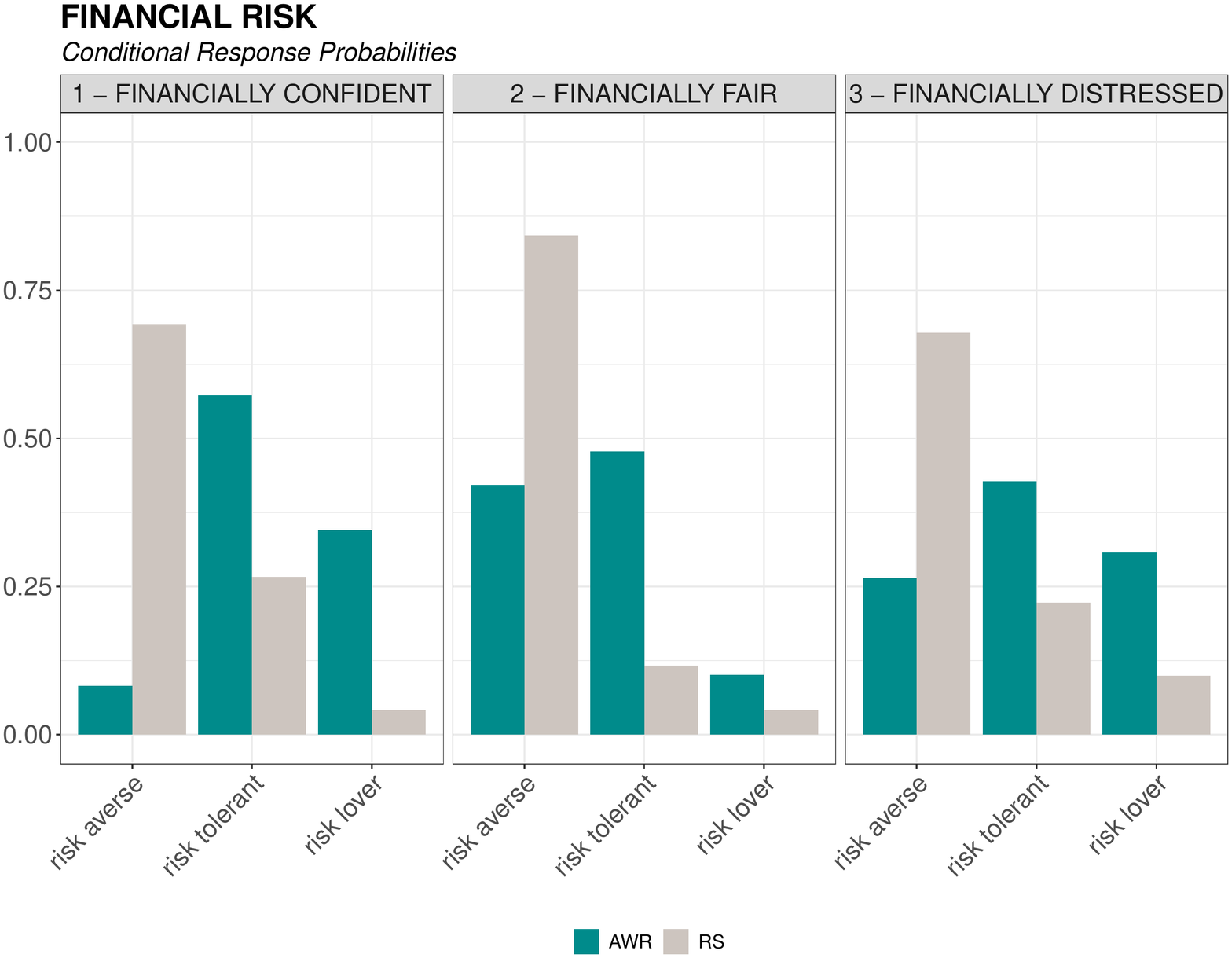}
	\captionsetup{width=.85\textwidth}
	\caption{Conditional response probability functions of AWR and  RS respondents in the three latent states of the financial condition}
	\label{ObsProb}
\end{figure}

\subsection{Model interpretation}
Figure~\ref{ObsProb} allows us to characterize the answers of AWR and RS respondents in the three latent states, for both response variables.
The top panels of Figure~\ref{ObsProb} illustrate the response probability functions of the perceived household's financial ability to make ends meet $R_1$ in the three stata of the latent construct for the AWR (colored bars) and RS (grey bars) regimes. According to Section~\ref{OBS}, the estimates $\hat{\phi}_{011}$ and $\hat{\phi}_{111}$ reported in Table~\ref{tablephi} imply that the probability function of RS respondents in the state $l=1$ has mode at the middle category ($c_1/2$) \textit{fairly easily}, at the middle point ($c_1/2 + 1$) \textit{fairly difficulty} in the state $l=2$, and at the extreme ($c_1$) \textit{very difficulty} in $l=3$, respectively. This means that individuals, in the group of financially safer households ($l=1$), when in doubt about their perceived capability, tend to choose with more chance the middle category (MRS) \textit{fairly easily}  on the optimistic side of the scale, the uncertain households with a fair capability ($l=2$) instead take refuge in the middle category (MRS) \textit{fairly difficulty} on the pessimistic side and, finally, the reluctant households in the group that deals with financial stress ($l=3$) show a tendency towards the worrying categories (DRS), with a remarkable preference for the extreme difficulty.
On the other hand, looking at the probability functions of AWR respondents to question $R_1$ we deduce that, in the latent class of financial confident families, aware people seem more optimistic than the RS respondents in the same latent group, and concentrate quite all the probabilities on the right categories meaning easy affordability, with mode at \textit{fairly easily}. In the intermediate state, two highly preferred middle points \textit{fairly easily} and \textit{fairly difficulty} characterize the  probability function of the AWR respondents who deal with a fair financial capability, while the RS respondents in the same group prefer more negative positions. The two most selected categories  move to \textit{fairly difficulty} and \textit{difficulty} for AWR people facing financial struggles ($l=3$). Thus, in this state, AWR respondents are prone to manifest their difficulties in managing family's financial resources, but are most of the RS respondents, who experience financial distress, extremely struggling to make ends meet.

Bottom panels in Figure~\ref{ObsProb} refer to the probability functions of the observed response $R_2$ about financial risk perception for AWR (colored bars) and RS (grey bars) households. The RS probability functions in every latent state have  mode at the smallest category \textit{risk averse} since all the parameters $\phi_{0l2}$ and $\phi_{1l2}$ are negatively estimated (Table~\ref{tablephi}) for $l=2,3$ and $\hat{\phi}_{012} < - \hat{\phi}_{112}$ in the first state. It seems that reticent respondents take refuge in the status of extreme risk-aversion, may be for not blaming themselves for their financial condition.  Completely different are the distributions of AWR respondents. We can clearly see left skewed, right skewed and quite symmetric probability functions, respectively in the group of financially confident ($l=1$), financially fair ($l=2$), and financially distressed ($l=3$) households, all with mode at the middle category \textit{risk tolerant}. The preference for \textit{risk averse} is more evident in the groups of more financially vulnerable (fair and distressed) households, who may find it prudent to avoid unnecessary or excessive financial risk. Instead, \textit{risk lovers} mainly belong to two categories of households: the ones confident with their financial plan and budget that can afford risky financial practices, and those respondents who got into financial difficulties because of their poor financial behaviour. Finally, \textit{risk averse} and \textit{risk tolerant} are the preferred responses of people with fair ability to manage their finance, thereby showing their propensity to stay out of financial troubles.

\begin{sidewaystable}[!htbp]
		\captionsetup{width=1\textwidth}
	\caption{Maximum Likelihood Estimates (MLE) of the parameters of the selected model M8 with $k=3$, and standard errors (\textit{SE}), $^*$ $95 \%$ confidence interval does not contain zero}
	\begin{center}\label{MLE}
		\resizebox*{0.9\textwidth}{!}{\renewcommand\arraystretch{1.2}
			%\hspace*{-2.3cm}
			\begin{tabular}{rlrccccccccccc} \cmidrule{4-14}
				&   &   &  \multicolumn{1}{|c|}{score} &       \multicolumn{3}{c|}{intercepts}       & \multicolumn{7}{c|}{covariates} \\
				\cmidrule{4-14}       &   &       &       &       &       &       & \multicolumn{1}{c}{G} & \multicolumn{1}{c}{Jse} & \multicolumn{1}{c}{Jhrs} & \multicolumn{1}{c}{CH} & \multicolumn{1}{c}{D} & \multicolumn{1}{c}{S} & \multicolumn{1}{c}{E} \\
				\cmidrule{1-14}    \multicolumn{1}{c}{\multirow{3}[2]{*}{$\pi^L_1$}} & parameters &       & $\mu_3$ & & $\alpha_{02}$ & $\alpha_{03}$ &  \multicolumn{7}{c}{$\boldsymbol{{\alpha}'_{1}}$} \\
				\cmidrule{4-14}
				& MLE   &       & $1.700^*$ & & $2.042^*$ & $3.140^*$ &  \multicolumn{1}{c}{0.224} & \multicolumn{1}{c}{$-1.045^*$} & \multicolumn{1}{c}{$-0.504^*$} & \multicolumn{1}{c}{-0.211} & \multicolumn{1}{c}{-0.043} & \multicolumn{1}{c}{$-1.150^*$} & \multicolumn{1}{c}{$-1.343^*$} \\
				& \textit{SE} &       & 0.223 &  & 0.370 & 0.383 & \multicolumn{1}{c}{0.139} & \multicolumn{1}{c}{0.244} & \multicolumn{1}{c}{0.148} & \multicolumn{1}{c}{0.134} & \multicolumn{1}{c}{0.155} & \multicolumn{1}{c}{0.239} & \multicolumn{1}{c}{0.208} \\ \midrule
				\multicolumn{1}{c}{\multirow{3}[1]{*}{$\pi^U_{1}$}} & parameters &  &     &       &       & $\bar{\alpha_0}$ & \multicolumn{7}{c}{$\boldsymbol{\bar{\alpha}'_{1}}$} \\\cmidrule{7-14}
				& MLE   &       &  &     &       & -0.201 & \multicolumn{1}{c}{$-0.390^*$} & \multicolumn{1}{c}{0.181} & \multicolumn{1}{c}{-0.138} & \multicolumn{1}{c}{-0.244} & \multicolumn{1}{c}{0.108} & \multicolumn{1}{c}{0.280} & \multicolumn{1}{c}{$0.739^*$} \\
				& \textit{SE} &       &   &    &       & 0.270 & \multicolumn{1}{c}{0.193} & \multicolumn{1}{c}{0.296} & \multicolumn{1}{c}{0.192} & \multicolumn{1}{c}{0.187} & \multicolumn{1}{c}{0.214} & \multicolumn{1}{c}{0.228} & \multicolumn{1}{c}{0.193} \\
				\midrule
				\multicolumn{1}{c}{\multirow{9}[1]{*}{$\pi^L_{l|\bar{l}}$}} & parameters & &      &       & $\beta_{021}$ & $\beta_{031}$ & \multicolumn{7}{c}{$\boldsymbol{{\beta}'_{11}}$} \\ \cmidrule{6-14}
				& MLE   &       &   &    & -0.699 & $-1.839^*$ & $1.448^*$ & \multicolumn{1}{c}{0.640} & \multicolumn{1}{c}{-0.362} & \multicolumn{1}{c}{$-0.954^*$} & \multicolumn{1}{c}{0.842} & \multicolumn{1}{c}{$-2.327^*$} & \multicolumn{1}{c}{$-1.652^*$} \\
				& \textit{SE} &     &  &       & 0.621 & 0.686 & \multicolumn{1}{c}{0.400} & \multicolumn{1}{c}{0.546} & \multicolumn{1}{c}{0.510} & \multicolumn{1}{c}{0.472} & \multicolumn{1}{c}{0.445} & \multicolumn{1}{c}{0.464} & \multicolumn{1}{c}{0.406} \\ \cmidrule{6-14}
				& parameters &       &  &     & $\beta_{012}$ & $\beta_{032}$ & \multicolumn{7}{c}{$\boldsymbol{{\beta}'_{12}}$} \\ \cmidrule{6-14}
				& MLE   &  &     &       & $-4.170^*$ & $-2.540^*$ & \multicolumn{1}{c}{0.905} & \multicolumn{1}{c}{$1.871^*$} & \multicolumn{1}{c}{-0.340} & \multicolumn{1}{c}{-0.283} & \multicolumn{1}{c}{0.041} & \multicolumn{1}{c}{$-2.583^*$} & \multicolumn{1}{c}{$1.248^*$} \\
				& \textit{SE} &    &   &       & 1.009 & 0.757 & \multicolumn{1}{c}{0.573} & \multicolumn{1}{c}{0.767} & \multicolumn{1}{c}{0.734} & \multicolumn{1}{c}{0.608} & \multicolumn{1}{c}{0.666} & \multicolumn{1}{c}{0.706} & \multicolumn{1}{c}{0.623} \\ \cmidrule{6-14}
				&   parameters    &   &    &       & $\beta_{013}$ & $\beta_{023}$ & \multicolumn{7}{c}{$\boldsymbol{{\beta}'_{13}}$} \\ \cmidrule{6-14}
				&  MLE     &    &   &       & $-14.042^*$ & $-11.960^*$ & \multicolumn{1}{c}{0.237} & \multicolumn{1}{c}{0.838} & \multicolumn{1}{c}{-0.021} & \multicolumn{1}{c}{0.361} & \multicolumn{1}{c}{0.297} & $9.481^*$ & \multicolumn{1}{c}{0.624} \\
				&  \textit{SE}     &   &    &       & 0.349 & 0.249 & \multicolumn{1}{c}{0.300} & \multicolumn{1}{c}{0.618} & \multicolumn{1}{c}{0.337} & \multicolumn{1}{c}{0.302} & \multicolumn{1}{c}{0.394} & \multicolumn{1}{c}{0.293} & \multicolumn{1}{c}{0.349} \\
				\midrule
				\multicolumn{1}{c}{\multirow{6}[0]{*}{$\pi^{U|L}(u|l,\bar{u})$}} & parameters & &       & $\bar{\beta}_{011}$ & $\bar{\beta}_{021}$ & $\bar{\beta}_{031}$ &       &       &       &       &       &       &  \\ \cmidrule{5-7}
				& MLE   &  &     & $-3.958^*$ & -5.590 & $-11.545^*$ &       &       &       &       &       &       &  \\
				& \textit{SE} &   &    & 0.972 & 3.791 & 0.023 &       &       &       &       &       &       &  \\ \cmidrule{5-7}
				& parameters &   &    & $\bar{\beta}_{012}$ & $\bar{\beta}_{022}$ & $\bar{\beta}_{032}$ &       &       &       &       &       &       &  \\ \cmidrule{5-7}
				& MLE   &    &   & $1.964^*$ & $2.563^*$ & $2.821^*$ &       &       &       &       &       &       &  \\
				& \textit{SE} &   &    & 0.231 & 0.334 & 0.338 &       &       &       &       &       &       &  \\
				\bottomrule
		\end{tabular}}
	\end{center}
\end{sidewaystable}
	%

%###########################################

Estimates of the parameters of the models for the initial and transition probabilities are in Table~\ref{MLE}. The reported standard errors are calculated using the OPIM method, even if all the  methods illustrated in Section~\ref{EM} and Appendix have been applied.  They provided quite overall similar results, also close to the standard errors obtained by the bootstrap method. Table~\ref{tab:SE} shows, for the sake of simplicity,  the standard errors of the estimators for the parameters of the models of the initial and transition probabilities of the latent construct, calculated with the three illustrated methods and the bootstrap technique. There is coherence in the results, except for some cases. Some numerical issues appear mostly in correspondence with high  estimates of  the parameters.
\begin{table}[h!]
	\centering
	\captionsetup{width=.85\textwidth}
	\caption{Standard errors (\textit{SE}) for the parameters of the models of the initial and transition probabilities of the latent construct, calculated with the methods illustrated in Section~\ref{EM} and Appendix: outer product information matrix (OPIM), observed information matrix (OIM), sandwich matrix (SDW), and bootstrap (BOOT)}
	\begin{center}
		\resizebox*{0.9\textwidth}{!}{\renewcommand\arraystretch{1.2}
			%\hspace*{-2.3cm}
			\begin{tabular}{clcccccccccc}
				\cmidrule{2-12}
				&   \multicolumn{1}{|c}{SE}   &  \multicolumn{1}{|c|}{score} &       \multicolumn{2}{c|}{intercepts}       & \multicolumn{7}{c|}{covariates} \\	\cmidrule{2-12}
				&          &             &     &       & \multicolumn{1}{c}{G} & \multicolumn{1}{c}{Jse} & \multicolumn{1}{c}{Jhrs} & \multicolumn{1}{c}{CH} & \multicolumn{1}{c}{D} & \multicolumn{1}{c}{S} & \multicolumn{1}{c}{E} \\\cmidrule{3-12}
				\multicolumn{1}{c}{\multirow{6}[2]{*}{$\pi^L_1$}} & &  $\mu_3$ & $\alpha_{02}$ & $\alpha_{03}$ &  \multicolumn{7}{c}{$\boldsymbol{{\alpha}'_{1}}$} \\
				\cmidrule{1-12}
				& OPIM & 0.2231 & 0.3702 & 0.3834 &  0.1385 & 0.2443 & 0.1480 & 0.1338 & 0.1549 & 0.2390 & 0.2079 \\
				&OIM & 0.2363 & 0.3763 & 0.3934 &  0.1401 & 0.2544 & 0.1532 & 0.1272 & 0.1495 & 0.2390 & 0.2212 \\
				&SDW & 0.3681 & 0.4848 & 0.5144 &  0.1640 & 0.4026 & 0.1985 & 0.1340 & 0.1980 & 0.2844 & 0.2935 \\
				& BOOT &0.2614 & 0.4190 & 0.4009 &  0.1417 & 0.2651 & 0.1684 & 0.1210 & 0.1391 & 0.2891 & 0.2133 \\
				\cmidrule{3-12}
				& &       & $\beta_{021}$ & $\beta_{031}$ & \multicolumn{7}{c}{$\boldsymbol{{\beta}'_{11}}$} \\ 	\cmidrule{1-12}
				\multirow{14}[0]{*}{$\pi^L({l|\bar{l}})$} & OPIM  & & 0.6210 & 0.6858 & 0.3996 & 0.5458 & 0.5100 & 0.4716 & 0.4454 & 0.4637 & 0.4057 \\
				& OIM    &  & 0.7240 & 0.7833 & 0.4408 & 0.5978 & 0.5328 & 0.4313 & 0.4984 & 0.4971 & 0.4432 \\
				& SDW   &  & 1.4397 & 1.5217 & 0.6823 & 0.7312 & 0.6020 & 0.5125 & 0.6133 & 0.7334 & 0.5986 \\
				& BOOT  & & 0.4424 & 0.4880 & 0.3771 & 0.3984 & 0.4504 & 0.3637 & 0.3242 & 0.3684 & 0.3490 \\
				\cmidrule{4-12}
				& &       &   $\beta_{012}$ & $\beta_{032}$ & \multicolumn{7}{c}{$\boldsymbol{{\beta}'_{12}}$} \\ \cmidrule{2-12}
				& OPIM  & & 1.0086 & 0.7573 & 0.5730 & 0.7669 & 0.7337 & 0.6080 & 0.6662 & 0.7058 & 0.6230 \\
				& OIM    & & 1.2990 & 0.8712 & 0.8269 & 0.9260 & 0.8681 & 0.7443 & 0.7004 & 1.1187 & 0.5891 \\
				& SDW   & & 2.7221 & 1.1781 & 1.7513 & 1.5455 & 1.2890 & 1.6226 & 0.9298 & 2.6344 & 0.8313 \\
				& BOOT  & & 0.8202 & 0.7802 & 0.5664 & 0.7465 & 0.6842 & 0.6489 & 0.7295 & 0.5843 & 0.5926 \\\cmidrule{4-12}
				&      &       & $\beta_{013}$ & $\beta_{023}$ & \multicolumn{7}{c}{$\boldsymbol{{\beta}'_{13}}$} \\ \cmidrule{2-12}
				& OPIM  & & 0.3486 & 0.2493 & 0.2999 & 0.6179 & 0.3370 & 0.3019 & 0.3942 & 0.2926 & 0.3486 \\
				& OIM   & & NA    & NA    & 0.2966 & 0.5482 & 0.3502 & 0.3083 & 0.3921 & NA    & 0.3454 \\
				& SDW  & & NA    & NA    & 0.4164 & 0.8712 & 0.4467 & 0.4439 & 0.4327 & NA    & 0.4096 \\
				& BOOT & & 0.8542 & 0.8273 & 0.2465 & 0.4785 & 0.3012 & 0.2644 & 0.3961 & 0.7738 & 0.2678 \\ \cmidrule{1-12}
		\end{tabular}}
		\label{tab:SE}%
	\end{center}
\end{table}%

By the sign of the estimates of the parameters of model (\ref{stereo_L1}), in Table~\ref{MLE} row 1, we deduce that at the first occasion employees, people without savings and with high education are in a worse financial status ($l=2,3$) with higher probability. This effect is strengthened for the status that describes greater financial incapability as the score $\hat{\mu}_3$ is greater than 1. In particular, for high educated people, the odds of being financially distressed  ($l=3$) instead of confident ($l=1$)  is quite 10 times the odds for low educated respondents. Similarly, for households with no savings (with an employee job), the odds of being in financial vulnerability ($l=3$) instead of being confident in managing the disposable income is 7 times (about 6 times) the odds when households can count on savings (on a self employee job). In addition, by considering the difference between the scores in discussing the effect of the educational level, it follows that $\frac{\pi^L_{i1}(3, E=over high school)}{\pi^L_{i1}(2, E=over high school )} = exp\{1.3434*(1.7-1)\} \frac{\pi^L_{i1}(3, E=up to secondary school)}{\pi^L_{i1}(2, E= up to secondary school)}$, at the first occasion, the propensity of strongly struggling to make ends meet ($l=3$) instead of managing their finances without much effort ($l=2$), for high educated respondents is about 2.5 times that of low educated ones.   Analogously, the odds ratios are 2.23 and 2.07 when groups of households with/without savings and with employee/self employee-householder, respectively, are compared.

Looking at the estimated parameters (row 4) of the RS initial probabilities modelled by (\ref{L2}), we deduce that at the beginning of the survey, female and low educated respondents seem more inclined towards response styles when describing their financial condition.

From the estimated parameters (rows 7, 10, 13, 15) of model (\ref{stereo L3}) with parallel restriction ($\nu_{l\bar l}=1$) for the transition probabilities of the latent financial capability,  it seems that, in two consecutive moments, women, highly educated, with no children and no savings with higher probability  move  from a financially confident  condition ($\bar{l}=1$) to a worse status of financial vulnerability ($l=2,3$). When the starting status corresponds to a fair financial confidence ($\bar{l}=2$), self-employees with no savings to rely on and a low level of education are more likely to move towards other levels of financial capability ($l=1,3$).
Individuals  who suffer in one occasion financial distress but can count on personal savings, tend to improve their condition in the next time by moving with more probability towards the stata of financial stability ($l=1,2$). Moreover, it is worthwhile to note that all the intercepts are negative, therefore there is evidence of a higher propensity to rest in the same previous financial status. This is more striking for households who experience financial distress ($\bar{l}=3$) and with very small probabilities pass to more comfortable conditions ($l=1,2$).

The estimated intercepts of model (\ref{L4}) for the RS transition probabilities (rows 18, 21), suggest that respondents tend to keep the same behavior in answering the two questions in two consecutive occasions, regardless the latent state which represents the current perceived financial capability.

\section{Concluding remarks} \label{concl}
 A HMM for longitudinal data of ordered categorical variables, that takes into account that responses can be determined by a RS, has been introduced. The proposed model is an extension of previous proposals both in the field of RS modeling and  of HMMs for longitudinal data. The new model  can cope with both RS effects and temporal dependence, but there are  some points that deserve further attention in future research.
Some open issues are: (i) testing time invariance against time dependence of the response  style component, (ii) the possibility of introducing more RS latent variables, specific to different sets  of a partition of the
response variables, in order to relax the assumption that, at a given time point $t$,  RS affects  all  response variables or none, (iii) the   introduction of covariate effects in the observation component.

Under assumption A2 and if the conditional  RS transition probabilities are homogeneous, the  hypothesis $\pi^{U|L}(u|l,\bar u)=d_{\bar{u}}(u)$ of time invariance of the RS indicator  constrains $2k$ parameters on the frontier of the parametric space. A  test based on the log likelihood ratio  statistic can be used but in this case the asymptotic distribution of the statistic is a mixture of chi-squared distributions known as chi bar squared distribution.
The test can be easily implemented as shown  in \cite{Barchi} and \cite{ColFor} who dealt with related problems.

 Point (ii) above can be based on the approach of   graphical HMMs by \cite{ColGio2015}. Regarding point (iii), Assumption B4 on the observation model can be relaxed by modelling the observation probabilities as function of individual covariates as an alternative to the presence of covariate effects on the latent component. This can be the case when the main interest is on the observed responses and the latent variable serves to account for time dependence and respondent's unobserved heterogeneity not explained by RS. %Further methodological developments are in this direction. %In the spirit of this paper this could be done by using stereotype logit models for the probability functions of AWR responses and by allowing the parameters $\phi_{1j}$ to be linear functions of covariates.

 %Assumption B4 of the observation model can be relaxed by modelling the observation probabilities as function of individual covariates.

\section*{Appendix: Standard errors}\label{SE}

We discuss three approaches to the estimation of the standard errors of the maximum likelihood estimator $\hat{\b \theta}$ of $\b \theta$. Hereafter, the upper index $(m)$,  $m\in\{L,U\}$,  will be omitted from  the vectors of covariates $\b x^{(m)}_i$ and $\b z_{it}^{(m)}$ to simplify the notation.

Let $\b y_i$, $i \in \mathcal I$, be a  realization of the $Tr$ observable  variables $Y_{jit}$, $j \in \mathcal R$, $t \in \mathcal T,$ collected in the vector $\b Y_i$.
The joint probability function of $\b Y_i$, conditioned on the vector of covariates $\b x_i, \b z_i$ ($\b z_i$ is obtained by stacking the $\b z_{it}, \: t>1$), is denoted by $q(\b y_i|\b x_i, \b z_i;\b \theta)$.
The log-likelihood function of the observations $\b y_i$, $i \in \mathcal I$, is:
\begin{equation*}
\ell(\b \theta)=\sum_{i=1}^n\log q(\b y_i|\b x_i, \b z_i;\b \theta),
\end{equation*}
and the vector of the score functions is:
\begin{equation*}\b s(\b \theta)=\sum_{i=1}^n\frac{\partial\log q(\b y_i|\b x_i, \b z_i;\b \theta)}{\partial \b \theta}=\sum_{i=1}^n\b s_i(\b \theta).\end{equation*}

The calculation of standard errors can be based on OIM, OPIM, SDW methods, as mentioned in Section~\ref{EM}.
We here sketch briefly some technical details of the three methods, an alternative approach is based on the well known parametric bootstrap technique.

The OIM can be computed using the Oakes identity \citep{Oakes1998}:
	\begin{equation*}
	\b J(\b \theta)=-\frac{\partial^{2} \ell(\b \theta)}{\partial \b \theta \partial \b \theta \tr}=-  \left[\frac{\partial^2Q(\b \theta|\bar{\b \theta)}}{\partial \b \theta \partial \b \theta \tr}_{|\bar{\b \theta}=\b \theta}+
	\frac{\partial^2Q(\b \theta|\bar{\b \theta)}}{\partial \bar{\b \theta} \partial \b \theta \tr}_{|\bar{\b \theta}=\b \theta}\right],
	\end{equation*}
	as shown by \cite{Bart2015}.
	The first term inside the square  brackets  is easy to compute, using the outputs of the last  M step, as it is block diagonal with blocks  given by the Hessian matrices of the six  addends of (\ref{l1}).
	The computation of the second term inside the square  brackets is more demanding as it requires the derivatives with respect to $\bar{\b \theta}$ of  $\log\delta_{it}^{(1)} (u,l;\bar{\b \theta})$ and $\log\delta_{it}^{(2)} (u,l;\bar u ,\bar l;\bar{\b \theta})$. These derivatives can be obtained as an output of the Baum-Welch forward-backward algorithm as described by \cite{Bart2015}. For every element $\bar \theta_h$ of $\bar{\b \theta}$, the terms of
	$-\frac{\partial^2Q(\b   \theta|\bar{\b \theta})}{\partial \bar \theta_h \partial \b \theta \tr}_{|\bar{\b \theta}=\b \theta}$ are obtained by the derivatives with respect to $\b \theta \tr $ of the six addends of (\ref{l1}) if $\delta_{it}^{(1)} (u,l;\bar{\b \theta})$ and  $\delta_{it}^{(2)} (u,l;\bar u ,\bar l;\bar{\b \theta})$ are  replaced by
	$\delta_{it}^{(1)} (u,l;\bar{\b \theta})\frac{\partial\log\delta_{it}^{(1)} (u,l;\bar{\b \theta})}{\partial\bar \theta_h }$ and
	$\delta_{it}^{(2)} (u,l;\bar u ,\bar l;\bar{\b \theta})\frac{\partial\log\delta_{it}^{(2)} (u,l;\bar u ,\bar l;\bar{\b \theta})}{\partial\bar \theta_h}$, respectively.

Notice that, when the necessary expected values and derivatives  are obtained from the Baum-Welch forward-backward algorithm, the computation of the standard errors require repeated calls to a function that estimates logit models. Matrix  $\b J(\b \theta)$ is estimated by $\hat{\b J}=\b J(\hat{\b \theta})$ where $\hat{\b \theta}$ is the MLE of $\b \theta.$ The standard errors of the maximum likelihood estimators are  estimated by the square roots of the diagonal elements of $\hat{\b J}^{-1}.$

 If the RS-HMM is correctly specified, estimates of the standard errors  can be also derived  from the OPIM matrix
$\b I(\b \theta)=\sum_{i}\b s_i(\b \theta)\b s_i(\b \theta)\tr$. The matrix $\b I(\b \theta)$  is  estimated by $\hat{\b I}=\b I(\hat{\b \theta})$
and  the estimated standard errors of the maximum likelihood estimators are the square roots of the diagonal elements of  $\hat{\b I}^{-1}.$
The matrix  $\hat{\b I}$  is easier to compute than $\hat{\b J}$, due to the  effort needed to compute $\frac{\partial^2Q(\b \theta|\bar{\b \theta)}}{\partial \bar{\b \theta} \partial \b \theta \tr}_{|\bar{\b \theta}=\b \theta}.$ 

Remind that the HMM with a RS component is misspecified if there does not exist a $\b \theta$ such that $\tau(\b y|\b x, \b z)=q(\b y|\b x, \b z;\b \theta)$ with probability 1 where, for every $\b x$, $\b z$, $\tau(\b y|\b x, \b z)$ is the \emph{true} probability function generating the  data.
In this case, $\hat{\b \theta}$ is a pseudo maximum likelihood estimator which is a consistent estimator of the pseudo-true value $\b \theta_0=argmin_{\b \theta}\left(E_{\b x, \b z}E_{\b y} \tau(\b y|\b x, \b z)\log\frac{\tau(\b y|\b x, \b z)}{q(\b y|\b x, \b z;\b \theta)}\right)$, see \cite{vuong1989} and \cite{white1982}. When the  RS-HMM is misspecified, estimated standard errors of the pseudo maximum likelihood estimators are given by the square roots of the diagonal elements of $\hat{\b J}^{-1}\hat{\b I}\hat{\b J}^{-1}.$
The  estimators of the standard errors, obtained in this way, are robust in the sense that they are consistent, independently from the correct specification of the model.
The matrix $n\hat{\b J}^{-1}\hat{\b I}\hat{\b J}^{-1}$ is a consistent estimator of the SDW matrix $\b A(\b \theta_0)^{-1}\b B(\b \theta_0)\b A(\b \theta_0)^{-1}$ where
$\b A(\b \theta_0)=-E_{\b x, \b z}E_y\frac{\partial^{2} \log q(\b y|\b x, \b z;\b \theta)}{\partial \b \theta \partial \b \theta \tr}$ and
$\b B(\b \theta_0)=E_{\b x, \b z}E_y\frac{\partial\log q(\b y|\b x, \b z;\b \theta)}{\partial \b \theta}\frac{\partial\log q(\b y|\b x, \b z;\b \theta)}{\partial \b \theta\tr}.$
The  SDW matrix  plays a central role in testing problems on misspecified models \citep{vuong1989}.
As all models are possibly misspecified, the estimator of the standard errors based on the sandwich matrix should be always used in practice. However,  computational complexity and numerical instability problems make the use of estimates based on  the matrix $\hat{\b I}$ more practical in the case of the model considered here.

\bibliographystyle{chicago}
\bibliography{hiddenbiblioFINAL}

\end{document}